\documentclass[aps, prb, twocolumn, floatfix, superscriptaddress, nofootinbib]{revtex4-2}
\usepackage{amsmath, amssymb, graphicx, epsfig, xcolor}
\usepackage[colorlinks=true, allcolors=blue]{hyperref}

\begin{document}

\title {Unveiling a Family of Dimerized Quantum Magnets in Ternary Metal Borides$^{\S}$}

\author{Zhen Zhang$^{\dag}$}
\address{Department of Physics and Astronomy, Iowa State University, Ames, IA 50011, USA}

\author{Andrew P. Porter$^{\dag}$}
\address{Department of Chemistry, Iowa State University, Ames, IA 50011, USA}
\address{Ames National Laboratory, U.S. Department of Energy, Ames, IA 50011, USA}

\author{Yang Sun$^{\ast}$}
\address{Department of Physics, Xiamen University, Xiamen 361005, China}

\author{Kirill D. Belashchenko$^{\ast}$}
\address{Department of Physics and Astronomy and Nebraska Center for Materials and Nanoscience, University of Nebraska-Lincoln, Lincoln, NE 68588, USA}

\author{Gayatri Viswanathan}
\address{Department of Chemistry, Iowa State University, Ames, IA 50011, USA}
\address{Ames National Laboratory, U.S. Department of Energy, Ames, IA 50011, USA}

\author{Arka Sarkar}
\address{Department of Chemistry, Iowa State University, Ames, IA 50011, USA}
\address{Ames National Laboratory, U.S. Department of Energy, Ames, IA 50011, USA}

\author{Kirill Kovnir$^{\ast}$}
\address{Department of Chemistry, Iowa State University, Ames, IA 50011, USA}
\address{Ames National Laboratory, U.S. Department of Energy, Ames, IA 50011, USA}

\author{Kai-Ming Ho}
\address{Department of Physics and Astronomy, Iowa State University, Ames, IA 50011, USA}

\author{Vladimir Antropov$^{\ast}$}
\address{Department of Physics and Astronomy, Iowa State University, Ames, IA 50011, USA}
\address{Ames National Laboratory, U.S. Department of Energy, Ames, IA 50011, USA}

\def\thefootnote{$\S$}\footnotetext{This document is the unedited Author's version of a Submitted Work that was subsequently accepted for publication in Journal of the American Chemical Society, copyright \textcopyright{} 2024 American Chemical Society after peer review. To access the final edited and published work see \url{https://pubs.acs.org/doi/10.1021/jacs.4c05478}.}\def\thefootnote{\arabic{footnote}}

\def\thefootnote{$\dag$}\footnotetext{Equal contribution.}\def\thefootnote{\arabic{footnote}}

\def\thefootnote{$\ast$}\footnotetext{Email: yangsun@xmu.edu.cn (Y.S.); belashchenko@unl.edu (K.D.B.); kovnir@iastate.edu (K.K.); antropov@iastate.edu (V.A.).}\def\thefootnote{\arabic{footnote}}

\date{\today}

\begin{abstract}
Dimerized quantum magnets are exotic crystalline
materials where Bose-Einstein condensation of magnetic excitations can
happen. However, known dimerized quantum magnets are limited to only a
few oxides and halides. Here, we unveil 9 dimerized quantum magnets and
11 conventional antiferromagnets in ternary metal borides
\emph{MT}B\textsubscript{4} (\emph{M} = Sc, Y, La, Ce, Lu, Mg, Ca, Al;
\emph{T} = V, Cr, Mn, Fe, Co, Ni). In this type of structure, 3\emph{d}
transition-metal atoms \emph{T} are arranged in dimers. Quantum
magnetism in these compounds is dominated by strong antiferromagnetic
interactions between Cr (both Cr and Mn for \emph{M} = Mg and Ca) atoms
within the structural dimers, with much weaker interactions between the
dimers. These systems are proposed to be close to a quantum critical
point between a disordered singlet spin-dimer phase, with a spin gap,
and the ordered conventional N\'{e}el antiferromagnetic phase. This new
family of dimerized quantum magnets greatly enriches the materials
inventory that allows investigations of the spin-gap phase. All the
quantum-, conventionally-, and non-magnetic systems identified, together
with experimental synthesis methods of a phase suitable for
characterization, provide a platform with abundant possibilities to tune
the magnetic exchange coupling by doping and study this unconventional
type of quantum phase transition. This work opens up new avenues for
studying the quantum magnetism of spin dimers in borides and establishes
a theoretical workflow for future searches for dimerized quantum magnets
in other families or types of materials.
\end{abstract}

\maketitle

\noindent \textbf{INTRODUCTION}

Quantum magnets have attracted great interest due to their many exotic
phenomena\textsuperscript{1,2}. In quantum magnets, tuning the quantum
phase transition across the quantum critical point (QCP) can lead to
different novel states of matter. One of the simplest kinds of quantum
magnets are those consisting of strongly coupled spin dimers.
Antiferromagnetic (AFM) intradimer coupling results in a disordered
quantum paramagnetic singlet ground state. Its first excited state is a
triplet state. Interdimer interactions make the triplet bands
dispersive\textsuperscript{3,4}. External parameters such as magnetic
field\textsuperscript{4--13}, pressure\textsuperscript{14--18}, and
doping\textsuperscript{19,20} have been shown to close the spin gap and
generate magnetic order. Various descriptions for the ground state of
the triplets, such as Bose-Einstein
condensate\textsuperscript{4,5,7--9,21}, a triplet
crystal\textsuperscript{6}, or a supersolid\textsuperscript{10}, have
been proposed. Bose-Einstein condensation (BEC) of the bosonic triplons
is a particularly interesting phenomenon\textsuperscript{2} since BEC is
well-known for its credit to the superconductivity of Cooper pairs and
the superfluidity of \textsuperscript{4}He. In dimerized quantum
magnets, at the QCP, the singlet ground state intersects the bottom of
the lowest triplet's band dispersion, where BEC
happens\textsuperscript{4}. Exploring novel quantum magnets and states
of matter in the vicinity of the QCP is of great significance.

So far, dimerized quantum magnets have been discovered, and the
associated BEC have been studied\textsuperscript{2} mostly in oxides and
halides such as
BaCuSi\textsubscript{2}O\textsubscript{6}\textsuperscript{4,8,9,22},
Ba\textsubscript{3}Mn\textsubscript{2}O\textsubscript{8}\textsuperscript{11--13,23},
and TlCuCl\textsubscript{3}\textsuperscript{5,7,14--19}. Quantum spin
dimers have yet been scarcely discovered and reported in boron
compounds. Boron compounds are versatile due to their unusual
electronic\textsuperscript{24}, magnetic\textsuperscript{24,25}, and
structural\textsuperscript{26} properties. For example, boron nitride
(BN) is a chemical analog to carbon; both are expected to adopt similar
one-, two-, and three-dimensional structures\textsuperscript{27}.
Borides are often reported to be suitable for high-temperature
ferromagnets\textsuperscript{28--31}, catalysis\textsuperscript{32--34},
thermoelectricity\textsuperscript{35}, and superhard
materials\textsuperscript{36--39}. Superconductivity has been discovered
in magnesium diboride (MgB\textsubscript{2})\textsuperscript{40--43}.
Nd\textsubscript{2}Fe\textsubscript{14}B is one of today's
best-performing permanent magnets\textsuperscript{44}. The formation of
binary, ternary, and multinary borides among boron and metal elements
provides a rich materials family to explore.

\begin{figure*}[t]
	\includegraphics[width=\linewidth]{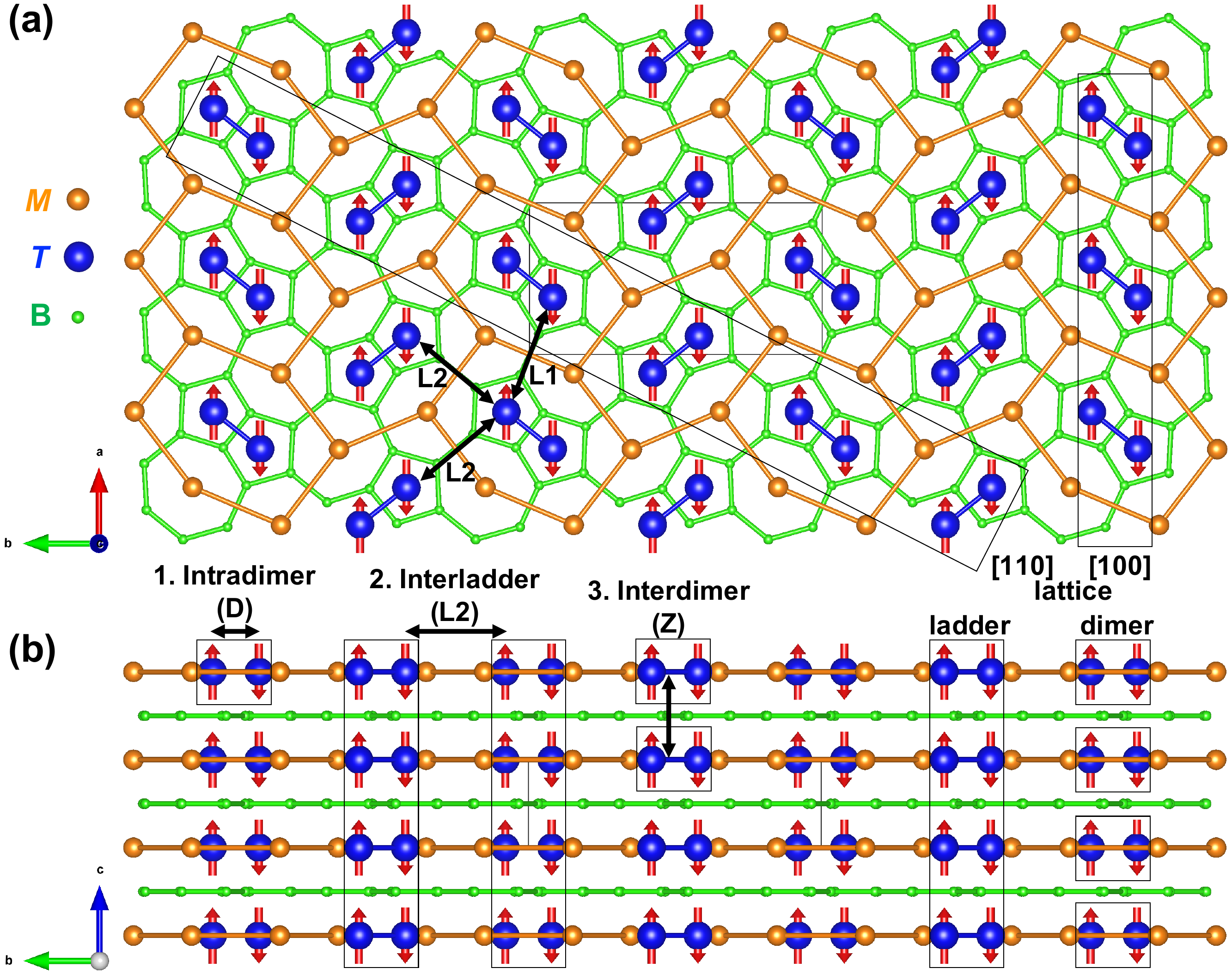}
	\caption{YCrB\textsubscript{4}-type crystal structure. (a)
		{[}001{]} and (b) {[}100{]} projections. Orange, blue, and green spheres
		show \emph{M}, \emph{T}, and B atoms, respectively. Red arrows show
		magnetic moments. AFF magnetic ordering is displayed in the figure as an
		example. The three-letter notation for the magnetic configuration
		indicates the relative alignment of the local moments inside a dimer
		(D), between the nearby dimers in the same plane (L2), and between the
		neighboring layers (Z).}
	\label{fig1}
\end{figure*}

Driven by the idea to seek quantum spin dimers in borides, we uncover
the family of compounds with YCrB\textsubscript{4}-type
structure. YCrB\textsubscript{4} was reported to offer promising
physical properties such as thermoelectricity\textsuperscript{45,46} and
high mechanical strength\textsuperscript{47--49}. This class of
materials, i.e., metal (\emph{M}) - transition metal (\emph{T}) -
tetraborides (B\textsubscript{4}), was reported by early experimental
studies\textsuperscript{50,51}. Sc\emph{T}B\textsubscript{4} (\emph{T} =
Fe, Co, Ni), Y\emph{T}B\textsubscript{4} (\emph{T} = V, Cr, Mn, Fe, Co),
Ce\emph{T}B\textsubscript{4} (\emph{T} = Cr, Mn, Fe, Co), and
Lu\emph{T}B\textsubscript{4} (\emph{T} = Cr, Fe, Co, Ni) were briefly
mentioned\textsuperscript{50--54}. However, methods to produce
single-phase samples suitable for characterization are known only for a
subset of the reported compounds. On the other hand, theoretical
investigations of the fundamental electronic and magnetic properties of
these compounds are scarce, even for the prototypical
YCrB\textsubscript{4}\textsuperscript{46,47,55}. Previous
first-principles calculations for YCrB\textsubscript{4} show
non-magnetic\textsuperscript{46,47} (NM) and
semiconducting\textsuperscript{46,47,55} behavior. A remarkable
structural feature of this class of materials is a transition-metal
dimer formation. How such a transition-metal dimer formation quantum
mechanically impacts the materials' electronic and magnetic properties
is unknown and to be explored.

\begin{figure*}[t]
	\includegraphics[width=\linewidth]{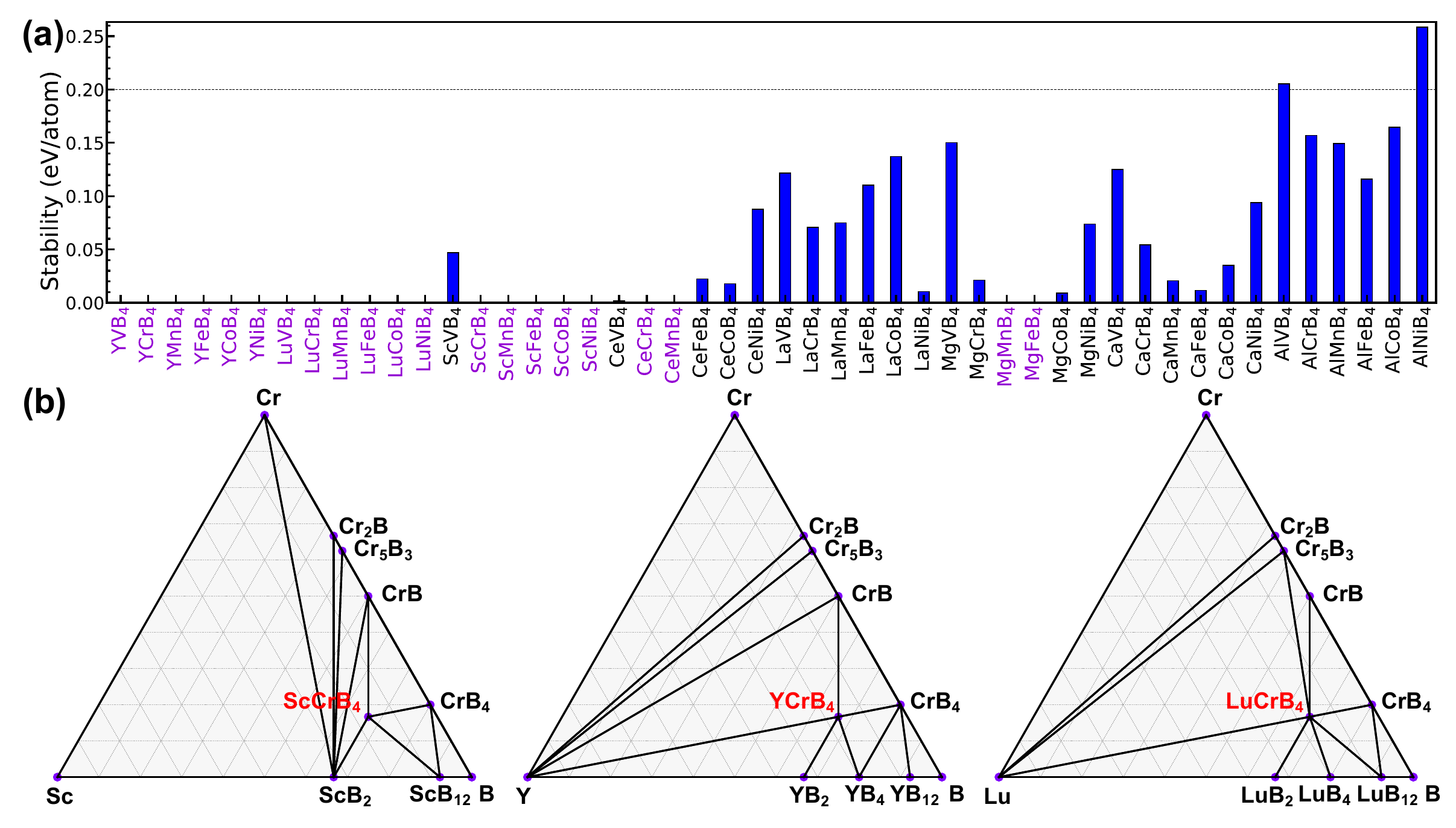}
	\caption{Phase stability. (a) Stability of
		\emph{MT}B\textsubscript{4}. Purple x-labels indicate stable compounds.
		(b) Convex hulls for \emph{M}-Cr-B (\emph{M} = Sc, Y, and Lu).}
	\label{fig2}
\end{figure*}

In this study, we utilize first-principles calculations to study the
electronic and magnetic properties of \emph{MT}B\textsubscript{4}, where
\emph{M} is IIIB metals (Sc, Y, La, Ce, Lu) and \emph{T} is 3\emph{d}
transition metals (V, Cr, Mn, Fe, Co, Ni). The NM IIIB elements are
considered to eliminate the magnetic effects arising from the M atoms on
the transition-metal dimers. Sc\textsuperscript{3+},
Y\textsuperscript{3+}, La\textsuperscript{3+}, and
Lu\textsuperscript{3+} all have a close-valence shell structure; thus,
their formed \emph{MT}B\textsubscript{4} compounds are expected to show
similar physical properties. Our previous study\textsuperscript{56}
generalized \emph{M} in this class of materials to the main-group
elements, Mg, Ca, and Al. However, whether they are quantum magnets
remains to be studied. In this work, we investigate the quantum
magnetism for this entire generalized family of materials.

From the synthesis point of view, YCrB\textsubscript{4} is the compound
that gives the name to this structure type. YCrB\textsubscript{4} and
YMnB\textsubscript{4} have been experimentally synthesized in
single-phase form, and their crystal structures were reported together
with some conventional properties\textsuperscript{45,46,49--51}. In
turn, YFeB\textsubscript{4} and YCoB\textsubscript{4} have been only
briefly mentioned without properly describing the experimental synthetic
procedures. In this study, experimental efforts are focused on the
synthesis of YFeB\textsubscript{4} and YCoB\textsubscript{4} and the
characterization of their structure and basic properties. Synthetic and
computational approaches were developed in parallel in this work to
further future studies of
abundant quaternary borides containing two IIIB metals or two 3\emph{d}
metals, i.e., doping on the \emph{M} site or the \emph{T} site.\\

\noindent \textbf{RESULTS AND DISCUSSION}

\textbf{YCrB\textsubscript{4}-Type Crystal Structure.} The
YCrB\textsubscript{4}-type crystal structure is displayed in Figure 1.
It crystallizes in the \emph{Pbam} space group with a layered structure,
consisting of alternating sheets of boron atoms and metal atoms. The
sheet of boron atoms forms a tiling of pentagons and heptagons. The
sheet of metal atoms consists of a tiling of squashed hexagons of
\emph{M} atoms with a transition-metal dimer inside each hexagon. Take
Y\emph{T}B\textsubscript{4}, for example. The intradimer distances (D)
are $\sim$2.3--2.6 \(\mathring{\mathrm{A}}\) for different
\emph{T} elements. Dimers form ladders in the direction perpendicular to
the plane ({[}001{]}). Dimer-dimer separation along the ladder (Z) is
$\sim$1.4 D. Ladders make up a two-dimensional lattice in
directions parallel to the plane ({[}110{]}, {[}1\(\overline{1}\)0{]},
or {[}100{]}). Ladder-ladder separation (L1, L2) is $\sim$2 D.
(Each \emph{T} atom has one L1 and two L2 neighbors at very similar
distances.) Since the intradimer distance is the shortest among the
three, intradimer interaction should be the strongest interaction among
the \emph{T} atoms.

\begin{figure*}[t]
	\includegraphics[width=\linewidth]{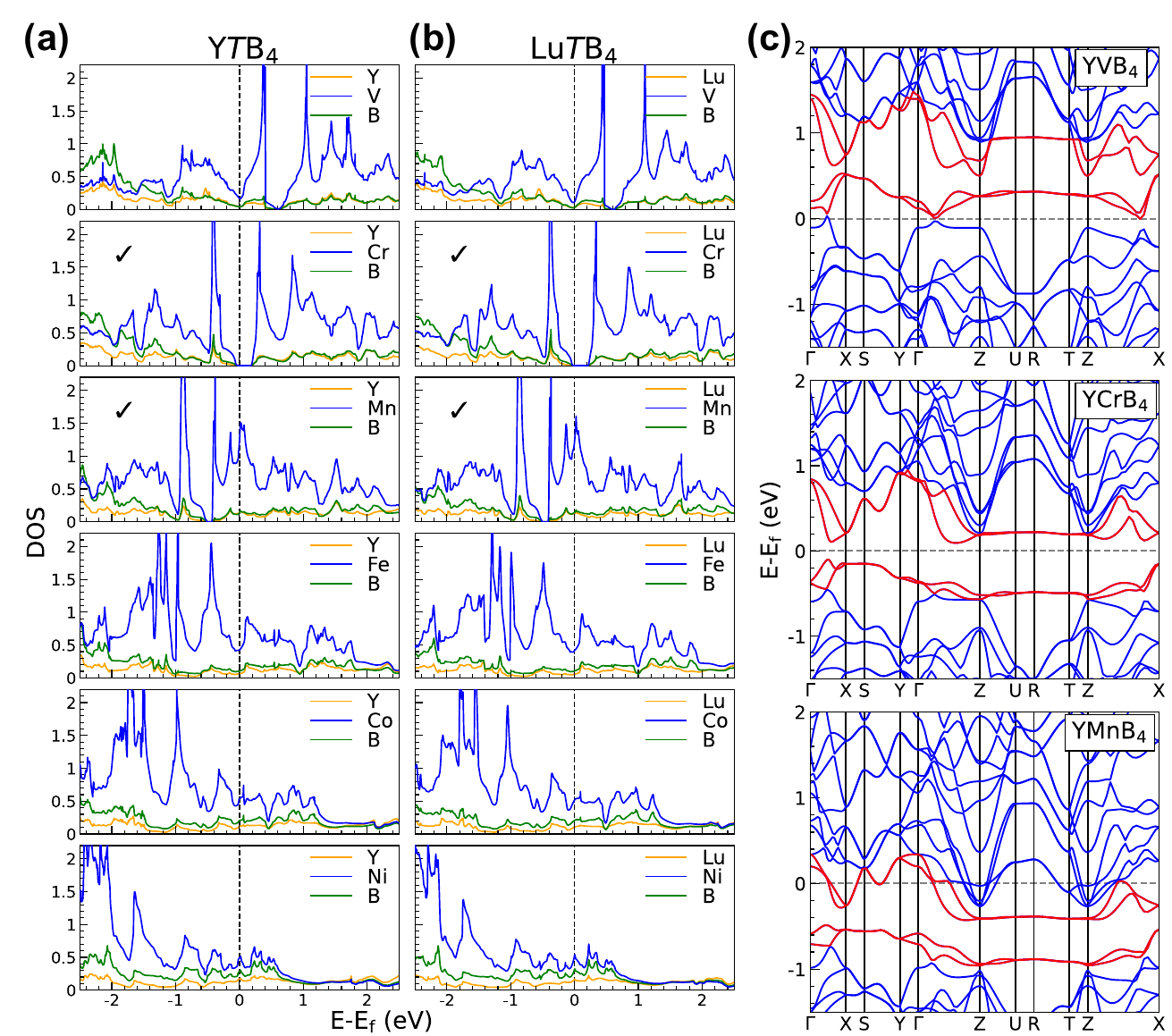}
	\caption{Non-magnetic (NM) electronic structures. Density of
		states (eV\textsuperscript{-1} f.u.\textsuperscript{-1}
		spin\textsuperscript{-1}) for (a) Y\emph{T}B\textsubscript{4} and (b)
		Lu\emph{T}B\textsubscript{4} (\emph{T} = V, Cr, Mn, Fe, Co, and Ni).
		Orange, blue, and green curves represent partial DOS of \emph{M},
		\emph{T}, and B atoms, respectively. Check marks indicate systems with
		stable magnetic solutions. (c) Band structure for
		Y\emph{T}B\textsubscript{4} (\emph{T} = V, Cr, and Mn). High-symmetry
		points in the Brillouin zone are denoted as follows: \(\Gamma\)(0, 0,
		0), X(1/2, 0, 0), S(1/2, 1/2, 0), Y(0, 1/2, 0), Z(0, 0, 1/2), U(1/2, 0,
		1/2), R(1/2, 1/2, 1/2), T(0, 1/2, 1/2). Red color highlights nearly flat
		bands in the Brillouin zone corresponding to two narrow peaks in the NM
		DOS.}
	\label{fig3}
\end{figure*}

\textbf{Phase Stability.} The formation energy (E\textsubscript{f}) is
obtained from spin-polarized calculations. Then, the formation energy
relative to the convex hull (E\textsubscript{d}) is evaluated by the
formation energy differences with respect to the three reference phases
forming the Gibbs triangle on the convex hull. The reference phases on
the convex hulls are obtained from materials databases such as Materials
Project\textsuperscript{57} and OQMD\textsuperscript{58}. All phases are
fully relaxed, and the total energies are calculated using the same
density-functional theory (DFT) settings. If an
\emph{MT}B\textsubscript{4} phase has E\textsubscript{d} \textless{} 0,
then it is stable. In this case, the convex hull is reconstructed to
include the \emph{MT}B\textsubscript{4}'s E\textsubscript{f}, and the
\emph{MT}B\textsubscript{4}'s stability is denoted as zero. If an
\emph{MT}B\textsubscript{4} phase has E\textsubscript{d} \textgreater{}
0, then it is not stable. In this case, the
\emph{MT}B\textsubscript{4}'s stability indicates the distance above the
convex hull. The calculated stabilities of all
\emph{MT}B\textsubscript{4} are shown in Figure 2a. Stability = 0
indicates a stable phase. We use 0.2 eV/atom\textsuperscript{59} as the
criterion for choosing metastable compounds that are possibly stabilized
by thermodynamics. Then, we obtain 21 stable and 25 metastable compounds
in \emph{MT}B\textsubscript{4}. The calculated convex hulls for
\emph{M}-Cr-B (\emph{M} = Sc, Y, and Lu) systems are exhibited in Figure
2b.

\textbf{Electronic Properties of the Non-Magnetic (NM) States.} The NM
density of states (DOS) for the representative Y and Lu compounds are
plotted in Figure 3, and the DOS at the Fermi level
N(E\textsubscript{f}) are listed in Table 1. Due to limited space, our
discussion will mainly focus on Y compounds. Other compounds have been
analyzed similarly. N(E\textsubscript{f}) of these compounds are mainly
contributed by the \emph{T} atom. V has a low DOS at the Fermi level,
making V compounds weakly metallic. Fe, Co, and Ni have
N(E\textsubscript{f}) $\sim$ 0.5 eV\textsuperscript{-1}
f.u.\textsuperscript{-1} spin\textsuperscript{-1}. Hence, these
compounds are ``good'' metals. For Fe, the Fermi level is at the minimum
of DOS, but there is a peak right above it. Applying some strain or
doping may move the Fermi level toward this peak. For Ni, the Fermi
level is located at a minor peak. In contrast, for Mn, the Fermi level
is at a sharp peak of DOS, suggesting a strong electronic instability.

\begin{table}[t]
	\caption{Density of states (DOS) of transition metals
		N\emph{\textsubscript{T}}(E\textsubscript{f}) (eV\textsuperscript{-1}
		f.u.\textsuperscript{-1} spin\textsuperscript{-1}) and total DOS
		N\textsubscript{tot}(E\textsubscript{f}) (eV\textsuperscript{-1}
		f.u.\textsuperscript{-1} spin\textsuperscript{-1}) at the Fermi level
		for Y\emph{T}B\textsubscript{4} and Lu\emph{T}B\textsubscript{4}.}
	\label{tab1}
	\begin{ruledtabular}
		\begin{tabular}{lcc|lcc}
			Compound & N\emph{\textsubscript{T}}(E\textsubscript{f}) & N\textsubscript{tot}(E\textsubscript{f}) & Compound & N\emph{\textsubscript{T}}(E\textsubscript{f}) & N\textsubscript{tot}(E\textsubscript{f}) \\
			\hline
			YVB\textsubscript{4} & 0.165 & 0.331 & LuVB\textsubscript{4} & 0.117 & 0.230 \\
			YCrB\textsubscript{4} & 0.0 & 0.0 & LuCrB\textsubscript{4} & 0.0 & 0.0 \\
			YMnB\textsubscript{4} & 1.209 & 1.696 & LuMnB\textsubscript{4} & 1.194 & 1.656 \\
			YFeB\textsubscript{4} & 0.425 & 0.680 & LuFeB\textsubscript{4} & 0.417 & 0.667 \\
			YCoB\textsubscript{4} & 0.537 & 1.016 & LuCoB\textsubscript{4} & 0.516 & 0.972 \\
			YNiB\textsubscript{4} & 0.488 & 1.142 & LuNiB\textsubscript{4} & 0.530 & 1.210 \\
		\end{tabular}
	\end{ruledtabular}
\end{table}

The NM DOS for YCrB\textsubscript{4} shows a semiconducting character
with a narrow band gap of 0.20 eV, which agrees with the experimental
estimate of 0.17 eV\textsuperscript{45}. The semiconducting character of
NM YCrB\textsubscript{4} agrees with previous first-principles
calculations\textsuperscript{46,47,55}, which reported a gap of 0.05
eV\textsuperscript{55}, 0.14$\pm$0.04 eV\textsuperscript{46}, and 0.17
eV\textsuperscript{47}.

A remarkable feature in the NM DOS of these compounds is the presence of
two narrow peaks, which are both empty in V, straddle the Fermi level in
Cr, and both filled in Mn compounds. These narrow peaks displayed in the
electronic
spectrum are known as Van Hove singularities (VHS). Generally, VHS near
the Fermi level can lead to instabilities, such as charge density waves,
and induce phase transitions under changing temperature, pressure, or
chemical doping. They can also be associated with interesting optical
properties and with large DOS near the Fermi level, the latter of which
tends to enhance superconductivity. Understanding the presence and
influence of VHS is valuable in the design and engineering of new
materials.

These VHS result from two pairs of bands, highlighted in red in Figure
3, that are nearly flat on the \(k_{z} = 1/2\) plane (Z-U-R-T-Z path in
Figure 3). The flat bands are formed by \(d_{z^{2}}\) orbitals of the
transition-metal atoms\textsuperscript{46}, and the two pairs of bands
correspond to bonding and antibonding states of the dimers. (Two bands
in each pair correspond to two dimers per unit cell.) For a generic
wavevector, the dimer states can hybridize with B atoms, leading to
significant dispersion visible along the $\Gamma$-X-S-Y-$\Gamma$-Z path. However, this
hybridization is suppressed at \(k_{z} = 1/2\) due to the destructive
interference between the hopping amplitudes for the two transition-metal
sheets to the \(s\), \(p_{x}\), and \(p_{y}\) orbitals of the B atoms.
This cancellation effectively localizes the dimer states with
\(k_{z} = 1/2\), leading to flat bands and the corresponding peaks in
the DOS. The strength of the interdimer hybridization determines the
width of the peaks and is, therefore, much smaller (at about 0.2 eV for
Cr compounds) than in Cr metal, where it is in the range of 1-2 eV.

The DOS for the different transition metals exhibits an approximately
``rigid band'' behavior. By tuning electron concentration and changing
the transition metal from V to Mn, the
DOS plots show similar patterns with the Fermi level located below, in
between, and above the two major VHS, respectively. Such a ``rigid
band'' behavior is common for similar systems\textsuperscript{56}.

\textbf{Magnetic States.} VHS near the Fermi level may be associated
with various electronic instabilities. Based on the analysis of the
real-space paramagnetic Pauli spin susceptibility matrix (see Methods),
we consider magnetic orderings with ferromagnetic (FM) or
antiferromagnetic (AFM) alignment of the local moments within a dimer,
between the nearby dimers in the same plane, and between the adjacent
layers (see Figure 1). We will use a three-letter notation consisting of
letters F (for FM) and A (for AFM) to denote these orderings, such as
AFF, where each letter identifies the alignments in this particular
order. The labeling of the first and the third letters is
straightforward. A few comments are in order about the in-plane
interladder ordering, which is indicated by the second letter. (1) It
indicates the ordering between dimers along the L2 but not L1 direction
(see Figure 1). (2) The alignment is identified by comparing spins
between DL2, which is the linked path of D and L2. (3) The A ordering is
similar to the stripe ordering on the hexagonal lattice. The specific
ordering patterns are displayed in Figure 4.

\begin{figure*}[t]
	\includegraphics[width=\linewidth]{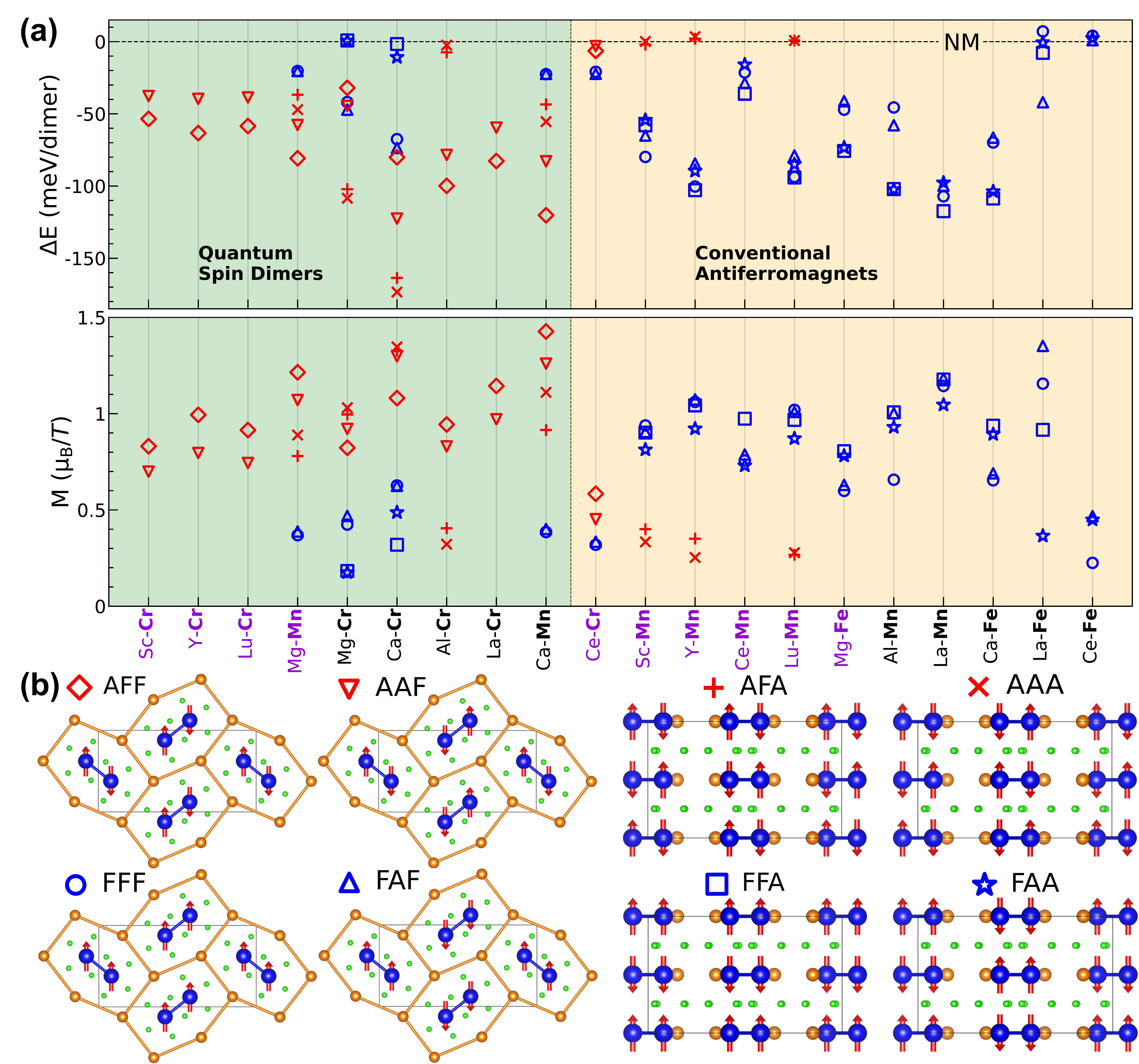}
	\caption{The DFT magnetic solutions. (a) The relative energy
		difference of the magnetic solutions to the NM one (top) and the
		magnetic moment on the transition metal \emph{T} (bottom) for
		\emph{MT}B\textsubscript{4}. The green and orange regions indicate
		quantum spin dimers and antiferromagnets, respectively. Purple and black
		x-labels indicate stable and metastable compounds, respectively. (b)
		Different magnetic configurations and the associated symbols and labels.}
	\label{fig4}
\end{figure*}

Eight magnetic configurations are obtained through all possible
combinations of these alignments and are considered in the calculations.
For example, AFF means intradimer AFM, interladder FM, and interlayer FM
coupling. The AFF configuration is displayed in Figure 1. However, not
all of these configurations are stable; many do not have a
self-consistent DFT solution with finite magnetic moments.

Figure 4a shows all the stable magnetic solutions for
\emph{MT}B\textsubscript{4}, with the eight magnetic configurations
illustrated in Figure 4b. The top panel of Figure 4a shows the
stabilization energy relative to the NM state, and the bottom panel
shows the magnetic moment on the \emph{T} atoms. Compounds without any
magnetic solutions are not shown. In total, we find 20 magnetic
compounds, among which 10 are stable (purple x-labels in Figure 4a) and
10 are metastable (black x-labels). Strong AFM intradimer exchange and
weak interdimer interactions result in quantum spin-gap systems.
According to the alignment of intradimer coupling of the DFT ground
state, we group these compounds into two categories: quantum spin-dimer
magnets and conventional magnets (see Figure 4). Such categorization
immediately justifies itself. It is clear that the DFT ground state
always has the same intradimer spin alignment, indicated by the symbol
color, as those of multiple low-energy states. Such a phenomenon results
from the dominating intradimer coupling strength compared to the
interdimer ones. Dimerized quantum magnets own AFM intradimer spin
alignments (red symbols in Figure 4) as DFT ground state and low-energy
states, while conventional magnets own FM intradimer spin alignments
(blue symbols) as ground and low-energy states.

\begin{figure*}[t]
	\includegraphics[width=\linewidth]{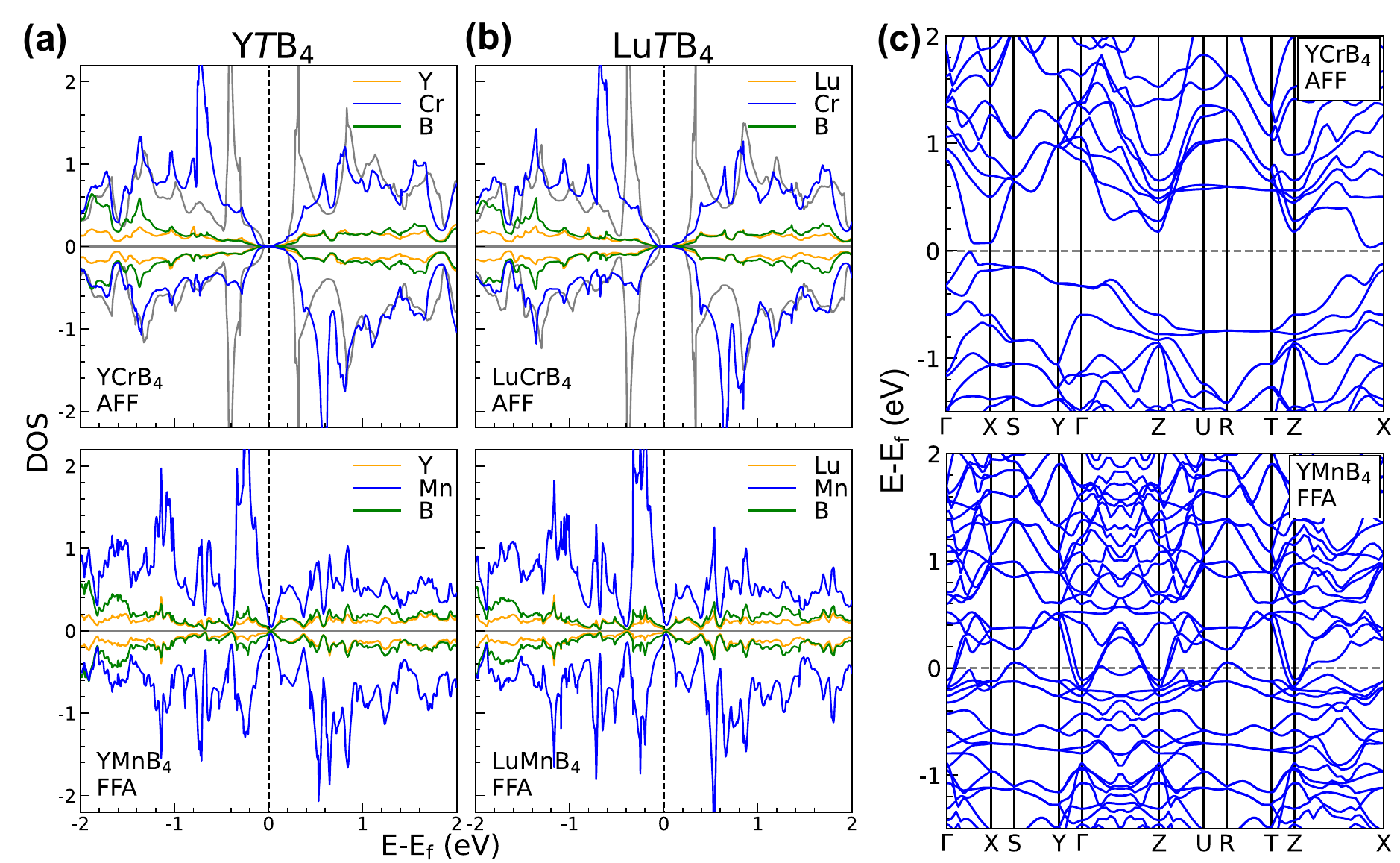}
	\caption{Magnetic electronic structures. The partial
		spin-polarized density of states (in eV\textsuperscript{-1}
		f.u.\textsuperscript{-1} spin\textsuperscript{-1}) of the DFT ground
		state of (a) Y\emph{T}B\textsubscript{4} and (b)
		Lu\emph{T}B\textsubscript{4}. \emph{T} = Cr has AFF solution, and
		\emph{T} = Mn has FFA solution. Positive (negative) DOS indicates
		majority (minority) spin. The corresponding NM partial DOS for a Cr atom
		is displayed as gray curves for comparison. (c) Spin-polarized
		electronic band structure for the magnetic ground-state
		YCrB\textsubscript{4} (top) and YMnB\textsubscript{4} (bottom).}
	\label{fig5}
\end{figure*}

We identify a total of 9 quantum spin-dimer magnets and 11 conventional
magnets. Note although the ground states of the conventional magnets
have FM intradimer ordering, they are still overall AFM with zero net
spins. The only exception is ScMnB\textsubscript{4}, the ground state of
which is overall FM (FFF). Considering the closeness of its energy to
other AFM states, we do not specifically distinguish this one and denote
all the conventional magnets as conventional antiferromagnets.
Remarkably, compounds with the same valence shell electron count belong
to the same magnets category and exhibit a similar distribution of
magnetic solutions. Cr forms a quantum AFM dimer with two ordered
solutions when combined with Sc\textsuperscript{3+},
Y\textsuperscript{3+}, La\textsuperscript{3+}, and
Lu\textsuperscript{3+}. Mn forms a conventional FM dimer with four
well-defined magnetic solutions when combined with
Sc\textsuperscript{3+}, Y\textsuperscript{3+}, La\textsuperscript{3+},
and Lu\textsuperscript{3+}. V, Fe, Co, and Ni are NM when combined with
these IIIB elements. Accordingly, the main-group element
Al\textsuperscript{3+} makes Cr quantum and Mn conventional with similar
distributions of magnetic solutions. Analogously, when it comes to
Mg\textsuperscript{2+} and Ca\textsuperscript{2+}, Mn and Fe play the
same role as Cr and Mn, respectively, for the 3+ elements. Therefore, Mn
forms quantum spin dimers when combined with Mg\textsuperscript{2+} and
Ca\textsuperscript{2+}, and Fe forms FM spin dimers with
Mg\textsuperscript{2+} and Ca\textsuperscript{2+}. In turn, Cr plays the
same role as V but is not as silent as V. Cr forms extra quantum spin
dimers when combined with Mg\textsuperscript{2+} and
Ca\textsuperscript{2+}. Ce case is more complicated due to the issue of
4\emph{f} electron localization. GGA+U may be required, which may change the
current results for Ce. We leave this to future study and only report
Ce's magnetism by GGA here.

Next, we analyze the magnetic solutions for the representative Y and Lu
compounds in detail. Earlier calculations predicted these systems to be
NM\textsuperscript{46,47}. However, we find the DFT ground states to be
AFF for Cr and FFA for Mn. In their respective DFT ground states, the
magnetic moments on Cr and Mn atoms are close to 1 \(\mu_{B}\). Besides
the ground states, self-consistent magnetic solutions were found in Cr
and Mn compounds with AAF ordering for Cr and FFF, FAF, FAA, AFA, and
AAA orderings for Mn. Among these, the AFA and AAA solutions for Mn have
small magnetic moments of about 0.3 \(\mu_{B}\) and the total energies
are very close to the NM solution. It may indicate a weak magnetic
moment dependence for the energy surface near the NM state, which could
enhance magnetic fluctuations.

Focusing on the solutions with well-defined magnetism,
YMnB\textsubscript{4} has stable magnetic solutions in all
configurations with FM intradimer alignment (FFF, FAF, FFA, FAA), which
all have similar energies (84--103 meV per dimer below the NM solution)
and local moments close to 1 \(\mu_{B}\). This reflects the small scale
of the interdimer exchange coupling compared to the intradimer coupling.
AFM interlayer orderings, FFA and FAA, are more stable than FM
interlayer orderings, FFF and FAF, by 2.5 and 5.2 meV per dimer,
respectively. FM interladder orderings, FFF and FFA, are more stable
than AFM interladder orderings, FAF and FAA, by 16 and 13 meV per dimer,
respectively. On the other hand, YCrB\textsubscript{4} has only two
stable magnetic solutions, AFF and AAF, which have AFM intradimer spin
alignment and energies 63 and 40 meV per dimer below the NM solution,
respectively. Interestingly, Cr only has AFF and AAF solutions,
indicating stable spin ladders with in-plane moment disorder with a
significant energy separation of 23 meV per dimer. This energy is like
that for YMnB\textsubscript{4} (16 and 13 meV per dimer) but slightly
larger. These observations for the Y compounds also apply to the Lu
compounds.

\begin{table*}[t]
	\caption{Exchange parameters \(J_{ij}\) (in mRy) in
		magnetic \emph{MT}B\textsubscript{4} (\emph{M} = Y and Mg).}
	\label{tab:table2}
	\begin{ruledtabular}
		\begin{tabular}{lccccccccccc}
			& & \multicolumn{2}{c}{AFF-YCrB\(_4\)} & \multicolumn{2}{c}{FFA-YMnB\(_4\)} & \multicolumn{2}{c}{AAA-MgCrB\(_4\)} & \multicolumn{2}{c}{AFF-MgMnB\(_4\)} & \multicolumn{2}{c}{FFA-MgFeB\(_4\)} \\
			\cline{3-4} \cline{5-6} \cline{7-8} \cline{9-10} \cline{11-12}
			Pair & \(N_{j}\) & \(R_{ij}\) & \(J_{ij}\) & \(R_{ij}\) & \(J_{ij}\) & \(R_{ij}\) & \(J_{ij}\) & \(R_{ij}\) & \(J_{ij}\) & \(R_{ij}\) & \(J_{ij}\) \\
			\hline
			D  & 1 & 2.36 & \textbf{-2.2}  & 2.44 & 2.00 & 2.35 & \textbf{-1.40} & 2.40 & \textbf{-1.36} & 2.42 & 0.94 \\
			Z  & 2 & 3.44 & 0.26  & 3.42 & \textbf{-0.09} & 3.25 & \textbf{-0.02} & 3.19 & 0.15  & 3.18 & \textbf{-0.15} \\
			DZ & 2 & 4.17 & \textbf{-0.40} & 4.20 & \textbf{0.08}  & 4.01 & 0.52  & 3.99 & \textbf{-0.17} & 4.00 & \textbf{-0.14} \\
			L1 & 1 & 4.83 & \textbf{-0.02} & 4.79 & 0     & 4.73 & \textbf{-0.03} & 4.70 & \textbf{0.004} & 4.67 & 0.02 \\
			L2 & 2 & 4.89 & \textbf{-0.03} & 4.81 & -0.01 & 4.79 & 0.03  & 4.73 & \textbf{-0.03} & 4.69 & -0.02 \\
			DL1 & 2 & 5.93 & 0     & 5.90 & 0.09  & 5.85 & -0.06 & 5.83 & 0.02  & 5.79 & 0.09 \\
			DL2 & 2 & 5.93 & -0.01 & 5.89 & 0.08  & 5.80 & \textbf{-0.06} & 5.76 & -0.03 & 5.73 & 0.04 \\
			ZL1 & 2 & 5.93 & \textbf{-0.15} & 5.88 & \textbf{-0.03} & 5.74 & -0.08 & 5.68 & \textbf{-0.11} & 5.65 & \textbf{0} \\
			ZL2 & 4 & 5.98 & \textbf{-0.11} & 5.90 & \textbf{-0.03} & 5.79 & \textbf{-0.06} & 5.71 & \textbf{-0.10} & 5.66 & \textbf{-0.01} \\
			\(J_{0}\) & & & 4.54 & & 2.59 & & 2.94 & & 2.49 & & 1.98 \\
			\(M\) &  & & 1.01 & & 0.91 & & 1.00 & & 0.93 & & 0.68 \\
			\(J_{D}/J_{0}\) & & & -0.48 & & 0.77 & & -0.48 & & -0.55 & & 0.47 \\
			\({|J}_{0}^{'}/J_{D}|\) & & & 1.06 & & 0.29 & & 1.10 & & 0.83 & & 1.11 \\
		\end{tabular}
	\end{ruledtabular}
\end{table*}

The three typical \emph{T}-\emph{T} distances for the NM and DFT
ground-state magnetic solutions are displayed in Figure S1 of the
Supporting Information. For the NM states, D, L2, and Z show
nonmonotonic trends with varying 3\emph{d} transition metals. For the
magnetic states of the Cr and Mn systems, Z is insensitive to magnetism,
whereas D is increased, and L2 is slightly
decreased by magnetism. D of both AFM Cr dimers and FM Mn dimers are
longer than their respective NM ones by $\sim$0.04 \AA. This
suggests that the intradimer exchange coupling in both Cr and Mn
compounds is strengthened when the intradimer bond length increases,
which is somewhat unusual. Note that the \emph{T}-\emph{T} distances for
different stable magnetic solutions agree with the DFT ground-state
ones, and those for the two marginally stable AFA and AAA solutions for
Mn are similar to those for the NM solution. Therefore, only DFT ground
states of the magnetic solutions are shown.

We now examine the electronic properties of the DFT ground states.
Figure 5 displays the spin-polarized DOS for the ground-state AFF Cr and
FFA Mn compounds. The results for the Y and Lu compounds are very
similar. AFF-ordered YCrB\textsubscript{4} is a magnetic semiconductor
with a small band gap of 0.05 eV. Compared to the NM partial DOS of a Cr
atom, which is shown in gray in Figure 5, the peaks in the magnetic DOS
are shifted by about 0.3 eV. FFA-YMnB\textsubscript{4} is a ``bad'' magnetic
metal with a low total DOS of 0.13 eV\textsuperscript{-1}
f.u.\textsuperscript{-1} spin\textsuperscript{-1} at the Fermi level.
While Mn atoms are aligned ferromagnetically in each layer, different
layers align antiferromagnetically along the z-direction. The
spin-polarized band structures for the ground-state
AFF-YCrB\textsubscript{4} and FFA-YMnB\textsubscript{4} are displayed
alongside the DOS in Figure 5.

The DFT ground states are either semiconducting or weakly metallic for
\emph{MT}B\textsubscript{4} (\emph{M} = Y and Lu). The semiconducting Cr
compounds in the NM state\textsuperscript{46,47,55} remain
semiconducting in the magnetic state. In contrast, a change from a good
metal to a bad metal is seen in Mn compounds when they switch from NM to
magnetic. Considering that conventional DFT functionals such as LDA and
GGA tend to underestimate the band gap, the band gaps in Cr compounds
are expected to be more significant. Mn compounds can also be magnetic
semiconductors.

Further, we calculate the exchange coupling parameters in the DFT ground
states of \emph{MT}B\textsubscript{4} using the linear response
technique\textsuperscript{60} implemented within the tight-binding
linear muffin-tin orbital (TB-LMTO) method in the Questaal
code\textsuperscript{61}. For YCrB\textsubscript{4}, we note that the
magnetic moment on the Cr atoms is sensitive to computational details
and is too small at 0.49 \(\mu_{B}\) in LMTO compared to the
full-potential VASP calculation (0.99 \(\mu_{B}\)). We correct this
discrepancy by scaling the exchange-correlation field by a factor of
1.13, which increases the magnetic moment to 1.01 \(\mu_{B}\).
Similarly, for MgCrB\textsubscript{4}, the exchange-correlation field is
scaled by 1.19 to obtain the local moment of 1.00 \(\mu_{B}\). For
YMnB\textsubscript{4}, MgMnB\textsubscript{4}, and
MgFeB\textsubscript{4}, the local magnetic moment in the ground state in
LMTO is 0.91, 0.93, and 0.68 \(\mu_{B}\), respectively.

The calculated exchange parameters are listed in Table 2. The parameters
are defined as the effective Heisenberg model,
\(E = - \sum_{ij}^{}J_{ij}{\widehat{m}}_{i}{\widehat{m}}_{j}\), where
each atomic pair is counted twice, and \({\widehat{m}}_{i}\) are unit
vectors showing the direction of magnetization for magnetic atom
\emph{i}. Positive (negative) exchange parameters correspond to FM (AFM)
coupling. Antiparallel pairs are highlighted in bold. \(N_{j}\): number
of neighbors of a given type.
\(R_{ij}\) (in \AA): distance to the given neighbor. D: dimer pair. Z:
nearest neighbor along the c axis. DZ: the nearest neighbor along the c
axis of the D neighbor. L1, L2: two inequivalent neighbor types between
the nearest ladders. A combination of symbols designates links of a path
to the other site.
\(J_{0} = \sum_{j}^{}J_{ij}{\widehat{m}}_{i}{\widehat{m}}_{j}\) is the
total exchange coupling between a given site and the rest of the
crystal. \(M\) (in \(\mu_{B}\)) is the local moment in LMTO.
\(J_{0}^{'}\) is similar to \(J_{0}\) but excludes the contribution from
the D neighbor.

The nearest-neighbor intradimer exchange coupling (denoted as D) is
dominant and quite strong in both YCrB\textsubscript{4} and YMnB\textsubscript{4}. This intradimer
exchange coupling is AFM in YCrB\textsubscript{4} and FM in YMnB\textsubscript{4}, consistent with the
magnetic ground states determined above based on total energy
calculations. The signs of all significant exchange parameters listed in
Table 2 for YCrB\textsubscript{4} agree with their relative alignment in
the ground state, indicating the absence of any magnetic frustration. In
YMnB\textsubscript{4}, only the weak DZ and L2 couplings are frustrated
among those listed in Table 2.

The last line in Table 2 lists the \({|J}_{0}^{'}/J_{D}|\) ratio, where
\(J_{0}^{'}\) is the total exchange coupling of a given magnetic moment
to the rest of the lattice, excluding its D neighbor. This ratio is 1.06
in YCrB\textsubscript{4} and only 0.29 in YMnB\textsubscript{4}. Because
the total coupling with other dimers is at most comparable with
intradimer coupling, the spins of the dimer partners should be strongly
correlated in the interesting temperature range where a magnetic phase
transition is possible. For YMnB\textsubscript{4} with FM intradimer
coupling and much weaker coupling to the rest of the lattice, it is
reasonable to assume that each dimer is maximally correlated, i.e.,
consider it a composite entity with effective spin 1. Because the
frustrated DZ and L2 couplings only amount to 15\% of \(J_{0}^{'}\), the
effect of this frustration on magnetic thermodynamics should be
negligible. Examination of the exchange parameters in
YMnB\textsubscript{4} also shows that the coupling has a
three-dimensional character; neither the interlayer nor interladder
coupling is anomalously small.

Because the exchange coupling between dimers in YMnB\textsubscript{4} is
three-dimensional without significant frustration, it is expected to
order antiferromagnetically at a temperature that may be estimated using
the mean-field approximation. The total coupling of a dimer to the rest
of the crystal is \(2J_{0}^{'}\). Treating the spins classically, we
obtain \(T_{N} \sim (2/3)2J_{0}^{'}/k_{B} \approx 120\) K.

Now consider YCrB\textsubscript{4}, where strong AFM intradimer coupling
favors the formation of a singlet spin-gap state with no magnetic
order\textsuperscript{62,63}. The singlet spin-gap state competes with
the antiferromagnetically ordered state favored by exchange coupling
between the dimers. Proceeding similarly to Ref.\textsuperscript{63}, we
consider a Heisenberg lattice spin-1/2 Hamiltonian where interaction
between dimers will be treated on the mean-field level, while the
\(4 \times 4\) Hamiltonian of a given dimer will be diagonalized
exactly. In this dimer Hamiltonian, interactions with other dimers are
represented by effective fields obtained by replacing the spin operators
for other dimers by their expectation values. The solution of this
quantum mean-field model predicts a QCP between the singlet spin-gap and
the magnetically ordered state at \({|J}_{0}^{'}/J_{D}| = 1\). With
\({|J}_{0}^{'}/J_{D}| = 1.06\), YCrB\textsubscript{4} is thus expected
to be close to the QCP.

Experimental measurements show that YCrB\textsubscript{4} has a very
small magnetic susceptibility, which was attributed to a small
concentration of magnetic impurities\textsuperscript{46}. In dimerized
spin-gap systems, the magnetic susceptibility is suppressed at
temperatures that are small compared to the singlet-triplet energy
gap\textsuperscript{64--66}. In the mean-field approximation for the
interaction between quantum spin-1/2 dimers, the susceptibility
is\textsuperscript{64--66}

\(\chi = \frac{k_{B}C}{k_{B}T\left( \frac{3}{4} + \frac{1}{4}\exp{(2\beta|J_{D}|}) \right) + {2J}_{F}^{'}}\), (1)

\noindent where it is assumed that \(J_{D} < 0\), \(C\) is the Curie-Weiss
constant, and \(J_{F}^{'} = \sum_{j}^{'}J_{ij}\) with \(J_{D}\)
excluded.

Using \(J_{D} = - 2.2\) mRy from Table 2, we can estimate the
singlet-triplet splitting for the isolated dimer Hamiltonian
\(H_{D} = - 2J_{D}{\widehat{\mathbf{\sigma}}}_{1}{\widehat{\mathbf{\sigma}}}_{2}\)
as \(2\left| J_{D} \right| = 4.4\) mRy. (We assume the DFT calculations
correspond to mean-field energies
\(E_{D} = - 2J_{D}{\widehat{m}}_{1}{\widehat{m}}_{2}\).) This
corresponds to a temperature of order 700 K, below which the magnetic
susceptibility should be suppressed due to the formation of a spin gap.

Exchange parameters for Mg compounds are also listed in Table 2. Mg has
a different valence shell electron count from Y. Two quantum spin-dimer
magnets, MgCrB\textsubscript{4} and MgMnB\textsubscript{4}, have
relatively weaker \(J_{D}\), -1.40 and -1.36 mRy, respectively, compared
to that of YCrB\textsubscript{4}. Thus, the spin gaps are expected to
cause a suppression for the magnetic susceptibility below a temperature
of around 450 K. Nevertheless, due to a relatively weaker \(J_{0}^{'}\),
MgCrB\textsubscript{4} and MgMnB\textsubscript{4} produce similar
\({|J}_{0}^{'}/J_{D}|\) ratios to that of YCrB\textsubscript{4}, 1.10
and 0.83, respectively, and thus are also expected to be close to the
QCP.

As for the conventional antiferromagnet, MgFeB\textsubscript{4}, the
frustrated L2 coupling only contributes to 2\% of \(J_{0}^{'}\). Hence,
the effect of frustration is weaker than the one for
YMnB\textsubscript{4}. On the other hand, the \({|J}_{0}^{'}/J_{D}|\)
ratio of MgFeB\textsubscript{4} is significantly larger than that of
YMnB\textsubscript{4} due to MgFeB\textsubscript{4}'s much weaker
intradimer coupling and stronger coupling of a dimer to the rest of the
crystal. We further estimate \(T_{N} \sim 220\) K, again, as in the case
of YMnB\textsubscript{4}, considering the dimer to be rigidly coupled.

By alloying the dimerized quantum magnets with different \emph{T}
elements on the \emph{T} sublattice or with different \emph{M} elements
on the \emph{M} sublattice, it may be possible to tune the exchange
coupling across the QCP, providing a rare platform and abundant
candidates for studying the spin-gap QCP.

\begin{figure*}[t]
	\includegraphics[width=\linewidth]{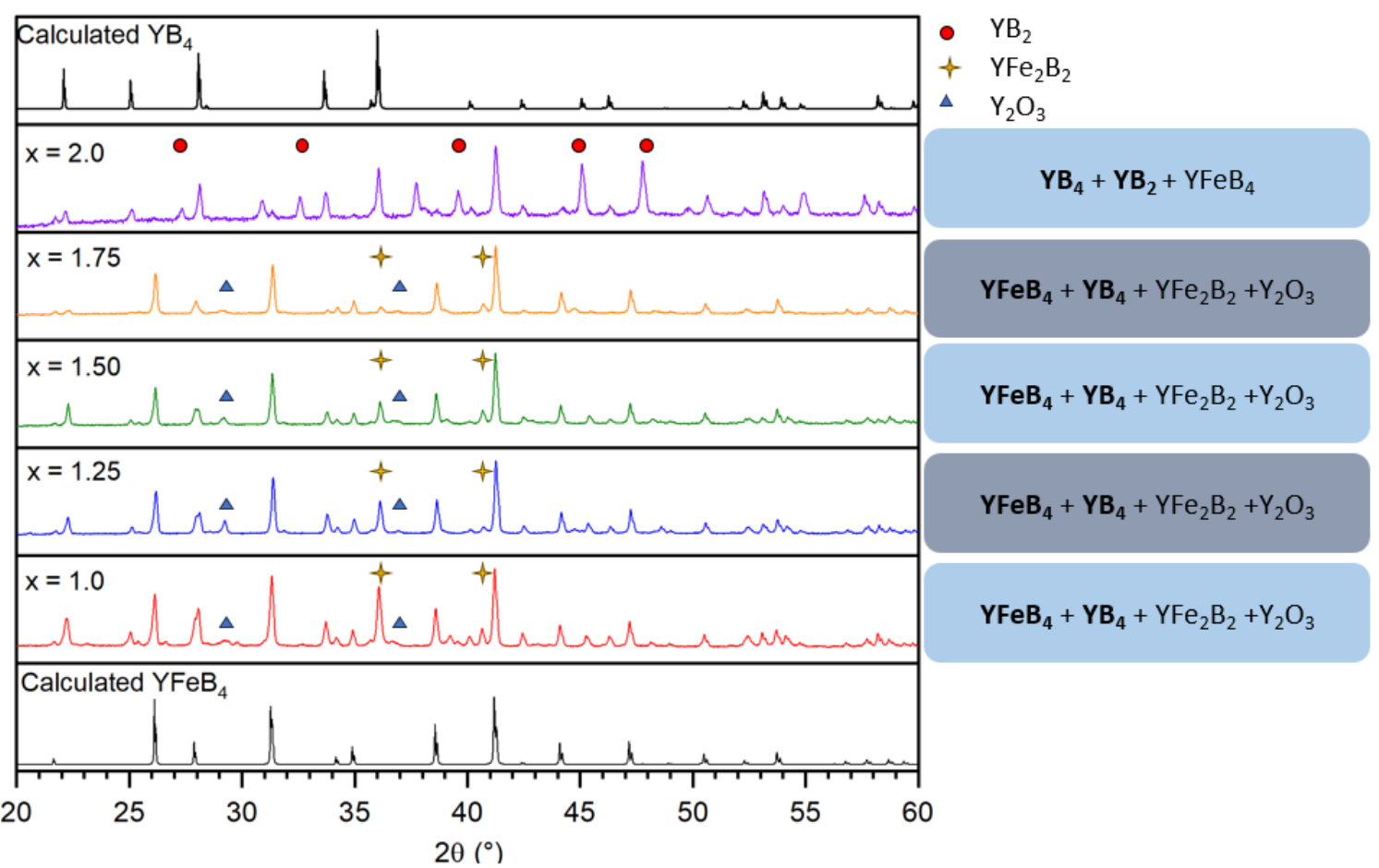}
	\caption{Experimental powder X-ray diffraction (PXRD) patterns
		(colored) of the arc-melted ingots with varying loaded composition of
		1Y+\emph{x}Fe+4B, where \emph{x} = 1.0, 1.25, 1.5, 1.75, and 2.0. The
		values of \emph{x} are indicated on the top left corner of the PXRD
		pattern. The calculated patterns of YFeB\textsubscript{4} (based on the
		single crystal structural model) and YB\textsubscript{4} are shown in
		black on the bottom and top, respectively. The phase composition for
		each sample, as determined by PXRD, is provided in the boxes to the
		right. The major phase(s) of each sample are indicated in bold. The most
		intense peaks of Y\textsubscript{2}O\textsubscript{3} (blue
		triangles), YFe\textsubscript{2}B\textsubscript{2} (yellow stars),
		and YB\textsubscript{2} (red circles) are also indicated.}
	\label{fig6}
\end{figure*}

\textbf{Experimental Synthesis.} The synthesis and basic properties of
YCrB\textsubscript{4} and YMnB\textsubscript{4} borides have already
been reported. In turn, for YFeB\textsubscript{4} and
YCoB\textsubscript{4}, the formation of the phases was mentioned, but
neither the procedure to produce single-phase samples nor properties
were reported\textsuperscript{45,46,49--51}. Our experimental interest
in Fe- and Co-containing compounds was two-fold: (i) to achieve future
targeted properties modifications by fine-tuning of the Fermi level
position via aliovalent substitution, the synthesis of corresponding
parent ternary phases need to be developed; (ii) such boron-rich layered
structures with boron networks analogous to MgB\textsubscript{2} call
for experimental verification of the presence or absence of
superconductivity in those compounds. We focused on the synthesis of
YFeB\textsubscript{4} and YCoB\textsubscript{4}. While the formation of
YCoB\textsubscript{4} was observed in the experimental samples as one of
the major phases, we were not able to reduce the amounts of admixtures
to a minimum level, and the properties of YCoB\textsubscript{4} were not
investigated.

\begin{figure*}[t]
	\includegraphics[width=\linewidth]{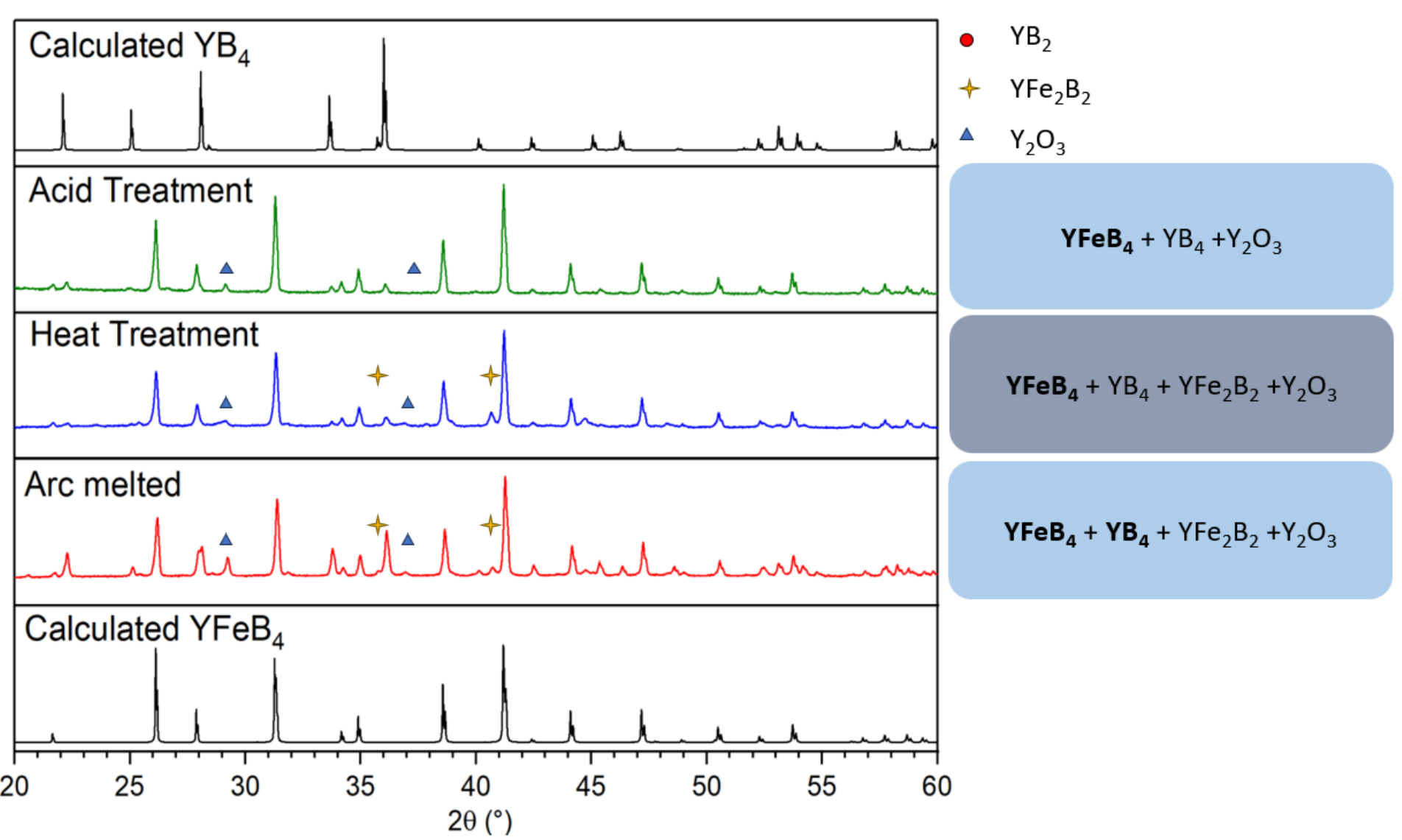}
	\caption{PXRD patterns of the sample with a loaded composition
		of Y+1.5Fe+4B. The sample was arc-melted (red pattern), then annealed
		isothermally at 1050 $^\circ$C for 5 days (blue pattern), followed by being
		ground into a fine powder and acid washed using 3M HCl (green pattern).
		Theoretical powder patterns of YB\textsubscript{4} and
		YFeB\textsubscript{4} (both black) are shown on top and bottom,
		respectively. The most intense peaks of
		Y\textsubscript{2}O\textsubscript{3} (blue triangles) and
		YFe\textsubscript{2}B\textsubscript{2} (yellow stars) are indicated.
		The boxes on the right reveal the phase composition of the samples. Bold
		text indicates a major phase.}
	\label{fig7}
\end{figure*}

Arc-melting of the stoichiometric YFeB\textsubscript{4} samples resulted
in ingots containing a mixture of two major phases,
YFeB\textsubscript{4} and YB\textsubscript{4}, along with minor
impurities of Y\textsubscript{2}O\textsubscript{3} and
YFe\textsubscript{2}B\textsubscript{2}. Single crystals of
YFeB\textsubscript{4} were obtained, and the crystal structure was
refined using single crystal X-ray diffraction (SCXRD). Synthetic
efforts were undertaken to characterize YFeB\textsubscript{4}
properties, aiming at a single-phase sample of YFeB\textsubscript{4}. To
suppress the formation of the YB\textsubscript{4} phase, an excess of Fe
was loaded
into the starting composition (see Figure 6). As expected, the excess of
Fe reduced the formation of YB\textsubscript{4} but resulted in an
incomplete elimination of the admixture. Moreover, samples with
\textgreater{} 25\% of Fe excess exhibited an increase in the formation
of YFe\textsubscript{2}B\textsubscript{2} impurity.

Further treatment was applied to reduce the impurity phases. Heat
treatment of the arc-melted ingots for 120 hours at 1050$^\circ$C was used to
improve the homogeneity of the sample. This allowed for
YB\textsubscript{4} and Fe-containing admixture phases to react further
and yield YFeB\textsubscript{4} (see Figure 7). Finally, an acid wash in
3M HCl was used to remove some of the remaining impurities, as the
target YFeB\textsubscript{4} phase is stable under these acid washing
conditions.

Thus, the optimized synthesis of YFeB\textsubscript{4} required the
arc-melting of samples containing an Fe excess, 25-50 at.\%. After
arc-melting, the samples were isothermally annealed at 1050$^\circ$C for 120
hours, followed by HCl acid washing the finely ground polycrystalline
samples overnight (see Figure 7). This optimized synthesis yields
YFeB\textsubscript{4} samples with minimal impurities, suitable for
basic properties characterization. Characterization of basic properties
for YFeB\textsubscript{4} is presented in Text S1 of the Supporting
Information.

\textbf{Single Crystal Diffraction.} The crystal structure of
YCrB\textsubscript{4} was recently redetermined with the single crystal
diffraction method\textsuperscript{67}. In general, in the structures of
borides, mixed occupancy and vacancies in metal sites may occur, which
can, in turn, affect the position of the Fermi level and result in
properties that are different from the expected properties for
stoichiometric compounds. We determined the crystal structure of
YFeB\textsubscript{4}. The refinement of the single crystal diffraction
data reveals 100\% occupancy of Fe and Y sites with no Y/Fe mixing or
vacancies. In the structure of YFeB\textsubscript{4} determined at 100
K, the B-B distances vary in the 1.689(9)-1.825(7) \AA{} range. This is
comparable to the range of B-B distances in YCrB\textsubscript{4}
determined at 294 K: 1.72-1.83 \AA{} range. Replacing Cr with Fe resulted in
a significant increase in the metal-metal distance in the dimer from
2.374 \AA{} for Cr-Cr to 2.539(1) \AA{} for Fe-Fe. The latter distance agrees with
the computationally predicted Fe-Fe distance (2.546 \AA, $\Delta$=0.28\%) for the
optimized structure. Further details of the crystal structure are
provided in Table 3 and in the Supporting Information as a
crystallographic information file (CIF).\\

\begin{table}[t]
	\caption{Data collection and structure refinement parameters for
		YFeB\textsubscript{4}. The deposition number 2325813 contains
		supplementary crystallographic data. These data are provided free of
		charge by the joint Cambridge Crystallographic Data Centre and FIZ
		Karlsruhe Access Structures service
		(\url{https://www.ccdc.cam.ac.uk/structures}).}
	\label{tab3}
	\begin{ruledtabular}
		\begin{tabular}{lc}
			\textbf{Compound} & YFeB\textsubscript{4} \\
			\hline
			CSD-numbers & 2325813 \\
			temperature (K) & 100 \\
			radiation (\AA) & Mo-\(K_{\alpha}\), 0.71037 \\
			crystal system & Orthorhombic \\
			space group & \(Pbam\) (no. 55) \\
			\(a\) [\AA] & 5.9049(3) \\
			\(b\) [\AA] & 11.4160(6) \\
			\(c\) [\AA] & 3.4094(2) \\
			Volume [\AA\(^3\)] & 229.83(2) \\
			\(Z\) & 4 \\
			data/parameters & 533/37 \\
			density (g/cm\(^3\)) & 5.43 \\
			\(u\) (mm\(^{-1}\)) & 31.01 \\
			\emph{R}\textsubscript{int} & 0.043 \\
			GOF & 1.38 \\
			\emph{R}\textsubscript{1}
			{[}\emph{I}~\textgreater~2$\sigma$(\emph{I}){]} & 0.028 \\
			\emph{wR}\textsubscript{2}
			{[}\emph{I}~\textgreater~2$\sigma$(\emph{I}){]} & 0.078 \\
			\emph{R}\textsubscript{1} & 0.030 \\
			\emph{wR}\textsubscript{2} & 0.080 \\
			diff. peaks [e \AA\(^{-3}\)] & 1.34/--1.31 \\
		\end{tabular}
	\end{ruledtabular}
\end{table}

\noindent \textbf{CONCLUSIONS}

In summary, using first-principles calculations, we identified 21
structurally stable and 25 metastable ternary metal borides of
\emph{MT}B\textsubscript{4}-type (\emph{M} = Sc, Y, La, Ce, Lu, Mg, Ca,
Al; \emph{T} = V, Cr, Mn, Fe, Co, Ni). Among them, we uncovered 20
magnetic systems, the rest being non-magnetic. Electronic and magnetic
calculations for this family of materials reveal similarities in their
electronic and magnetic structures for those compounds with the same
valence shell electron count. Magnetism in these compounds is dominated
by strong antiferromagnetic Cr (both Cr and Mn for \emph{M} = Mg and Ca)
or ferromagnetic Mn (Fe for \emph{M} = Mg and Ca) interactions within
the structural dimers with much weaker interactions between the
dimers. The magnetic ground states in DFT are semiconducting or weakly
metallic for these systems. Mn compounds are predicted to be
conventional N\'{e}el antiferromagnets with layered (A-type) ordering and
\(T_{N}\) below room temperature. In contrast, Cr compounds are proposed
to be close to a quantum critical point between a singlet spin-dimer
phase, with a spin gap above room temperature, and the conventional N\'{e}el
antiferromagnetic phase. In total, we unveiled 9 dimerized quantum
spin-gap systems and 11 conventional antiferromagnets. 4 of the
dimerized quantum magnets are stable (ScCrB\textsubscript{4},
YCrB\textsubscript{4}, LuCrB\textsubscript{4}, and
MgMnB\textsubscript{4}) and 5 are metastable (MgCrB\textsubscript{4},
CaCrB\textsubscript{4}, AlCrB\textsubscript{4}, LaCrB\textsubscript{4},
and CaMnB\textsubscript{4}). They provide a unique possibility for
investigating Bose-Einstein condensation of magnetic excitations in
crystalline systems\textsuperscript{2}. The prediction of this new
family of dimerized quantum magnets greatly expands the materials
inventory that allows such an investigation\textsuperscript{2}.
Experimental methods to produce single-phase YFeB\textsubscript{4}
suitable for characterization and YFeB\textsubscript{4}'s measured
crystal structure are reported. All the stable and metastable
\emph{MT}B\textsubscript{4} systems identified in this study, whether
quantum-magnetic, conventionally-magnetic, or non-magnetic, provide a
platform with numerous possibilities for future doping on the \emph{M}
site or the \emph{T} site. Tuning the magnetic exchange coupling by
doping may facilitate the study of this rare type of quantum phase
transition across the spin-gap quantum critical point. This work offers
new opportunities for studying the quantum magnetism of spin dimers in
borides. The step-by-step methodology used in this work also
demonstrates a theoretical workflow for future searches for dimerized
quantum magnets in other families or types of materials.\\

\noindent \textbf{METHODS}

\textbf{Density-Functional Theory Calculations.} We conducted DFT
calculations using the projector augmented wave (PAW)
method\textsuperscript{68} as implemented in the VASP
package\textsuperscript{69}. The exchange-correlation energy was treated
by the Perdew-Burke-Ernzerhof (PBE)\textsuperscript{70} generalized
gradient approximation (GGA). A plane-wave-basis set was used with a
kinetic energy cutoff of 520 eV. The convergence thresholds were
10\textsuperscript{-5} eV for electronic self-consistency and 0.01 eV
\AA\textsuperscript{-1} for ionic relaxation. The tetrahedron method with
Bl\"ochl corrections was used for the Brillouin zone integration. In the
structural optimization, the Brillouin zone was sampled by the
Monkhorst-Pack scheme with a k-point grid of 2$\pi$ $\times$ 0.033
\AA\textsuperscript{-1}. In the static calculations, energy, magnetic
moments, and electronic density of states were computed with a denser
k-point grid of 2$\pi$ $\times$ 0.022 \AA\textsuperscript{-1}.

\textbf{Real-Space Paramagnetic Pauli Spin Susceptibility Matrix
	Calculations.} Electronic instabilities suggest that analysis of the
possibility of local magnetic moments and magnetic order stabilization
is required. Take Y compounds, for example. As discussed in the main
text, Fe, Co, and Ni compounds are good metals, but according to the
Stoner criterion, their N(E\textsubscript{f}) is insufficient to
generate FM instability. In contrast, in YMnB\textsubscript{4}, the
Fermi level is located at a peak with N(E\textsubscript{f})
\textgreater{} 1 eV\textsuperscript{-1} f.u.\textsuperscript{-1}
spin\textsuperscript{-1}. Thus, ferromagnetism in Mn compounds is
plausible, while the insulating state of Cr compounds suggests
antiferromagnetism as the most likely option. Accordingly, we calculated
the real-space paramagnetic Pauli spin susceptibility matrix (see
details in Ref.\textsuperscript{71}) for transition metal atoms and
their neighbors. As expected, the important elements of \(\chi_{ij}\)
involve transition-metal atoms and their dimer partners. The on-site
term \(\chi_{00}\) dominates, and the intradimer matrix element
\(\chi_{ij}\) is about 10\% of the former in magnitude. The Anderson
local moment criterion (\(S_{00} = I\chi_{00} > 1\)) is barely satisfied
in Mn (\(S_{00}\) = 1.08), marginal in Cr (\(S_{00}\) = 0.98), and not
satisfied in all other systems. However, the generalization of the
Anderson local moment criterion is satisfied for a FM dimer in
YMnB\textsubscript{4} with \(I(\chi_{00} + \chi_{01})\) = 1.2 and for an
AFM dimer in YCrB\textsubscript{4} with \(I(\chi_{00} - \chi_{01})\) =
1.1. For all other systems, local moments on a single dimer are
unstable. Thus, while the local magnetic instability on Cr and Mn atoms
is likely, the resulting atomic moment can be classified as weakly
local. This can be compared to strong local moment systems such as bcc
iron, where \(S_{00}\) = 1.6\textsuperscript{71}. The values obtained
above are somewhat similar to the situation in hcp
cobalt\textsuperscript{71}. However, calculations of susceptibility in
cases of weak local moments may not be reliable, and only more precise
full-potential self-consistent calculations of the magnetic ground state
as seen in the main text can provide a justified answer.

\textbf{Synthesis.} High-purity Y fingers (99.9999 \%) were acquired
from the Materials Preparation Center at Ames National Laboratory. Other
chemicals were received from Alfa Aesar: Fe powder (99.9\%), Co powder
(99.5\%), and amorphous B powder (98\%). Y fingers were cut into small
pieces and weighed out to approximately 230 mg. Based on Y weight and
desired molar ratio, i.e., 1:1:4 or 1:1.5:4, Fe and amorphous B amounts
were recalculated and weighed out. The mixture of Fe and B was then cold
pressed into a pellet. The total sample weight (Y+Fe+B) was
approximately 500 mg. Yttrium pieces and the (Fe+4B) pellet were placed
onto a water-cooled copper hearth in a tungsten anode arc melter. The
chamber was evacuated for 20 minutes and filled with Ar gas.
Additionally, a Zr getter was placed in the chamber and arced to remove
any traces of oxygen. The YFeB\textsubscript{4} sample was arc-melted,
and then the ingot was flipped and re-melted. This process was repeated
four times. A similar procedure was applied for Co-containing samples.

\textbf{Annealing.} After arc melting, selected samples were loaded into
a carbonized silica ampoule, which was evacuated and flamed sealed. The
ampoules were placed into a muffle furnace and heated to 1050$^\circ$C with a
heating rate of 175$^\circ$C per hour. Once at 1050$^\circ$C, the samples were
isothermally annealed for 120 h. The furnace was turned off and allowed
to cool. After annealing, the ingot sample was ground into a fine powder
using a tungsten carbide mortar in air.

Selected annealed samples were acid-washed to reduce impurities. Our
studies indicated that the target YFeB\textsubscript{4} phase is
acid-stable. The samples were washed in 3M hydrochloric acid overnight
to remove YB\textsubscript{4} and any Fe impurities. After that, the
sample was filtered, washed with distilled water, and dried on air.

\textbf{Powder X-ray Diffraction (PXRD).} The samples were characterized
using powder X-ray diffraction (PXRD) on a Rigaku MiniFlex 600 powder
diffractometer with Cu-\emph{K}\textsubscript{a} radiation ($\lambda$ =
1.54059 \AA) and a Ni-\emph{K}\textsubscript{b} filter. Fluorescence
reduction was used to minimize background fluorescence from Fe.

\textbf{Single Crystal X-ray Diffraction (SCXRD).} Single crystals of
YFeB\textsubscript{4} were picked out from a crushed ingot with a loaded
composition of Y+Fe+4B. No further treatment was done to the ingot after
arc melting. Single crystal diffraction experiment was conducted using a
Bruker D8 Venture diffractometer with a PHOTON detector and
Mo-\emph{K}\textsubscript{a} radiation. The crystal was cooled to 100 K
with a stream of nitrogen gas. Structure determination and the final
refinement of the dataset were carried out using the SHELX suite of
programs. Refining of the occupancies of Fe and Y indicated no
deviations from 100\% within one standard uncertainty.

\textbf{Scanning Electron Microscopy and Energy Dispersion X-ray
	Spectroscopy (SEM EDS).} Elemental analysis was carried out using an FEI
Quanta 250 field emission SEM with an EDS detector (Oxford X--Max 80,
ThermoFischer Scientific, Inc.). The resulting data were analyzed using
the Aztec software. Powdered samples were mounted in epoxy, polished to
a level surface, and coated with a conductive layer of carbon. An
accelerating voltage of 15 kV was used. Compositions were normalized
with respect to 1 Y atom to provide information about the Y:Fe ratio.
The quantification of light elements, such as B, is not accurate using
this technique, but the presence of B was evident from EDS spectra and
elemental mapping (see Figure S4 of the Supporting Information).

\textbf{Magnetic Measurements.} Magnetic properties were measured by an
MPMS-XL SQUID magnetometer (Quantum Design) on two polycrystalline
YFeB\textsubscript{4} samples. The temperature dependence of the
magnetic susceptibility was measured from 2 K to 300~K at applied
magnetic fields of 0.1~T and 2T. The isothermal magnetization field
dependence was measured from --7 T to 7~T at 2~K and 300 K.

\textbf{Resistivity Measurements.} A physical property measurement
system (PPMS, Quantum Design) was used to characterize the electrical
resistivity of the arc-melted ingot of YFeB\textsubscript{4} (see Figure
S5 of the Supporting Information). The ingot was sanded down until a
smooth flat surface remained. 50 mm platinum wire was attached to the
smooth surface using conductive silver paste. A four-probe method was
used for the measurements.

\textbf{Heat Capacity Measurements.} Heat capacity was measured from
3.5-300 K using the heat capacity option of the Quantum Design Physical
Property Measurement System. The small part of the ingot (13.4 mg) was
mounted to the sample platform with Apiezon ``N'' grease. An addenda
measurement containing the heat capacity of the grease was subtracted
from the total heat capacity.\\

\noindent \textbf{ACKNOWLEDGMENT}

This work is supported by the U.S.
Department of Energy (DOE) Established Program to Stimulate Competitive
Research (EPSCoR) Grant No. DE-SC0024284. Computations were performed at
the High Performance Computing facility at Iowa State University and the
Holland Computing Center at the University of Nebraska.\\

\noindent \textbf{REFERENCES}

(1) Balents, L. Spin Liquids in Frustrated Magnets. \emph{Nature}
\textbf{2010}, \emph{464}, 199--208.
\url{https://doi.org/10.1038/nature08917}.

(2) Zapf, V.; Jaime, M.; Batista, C. D. Bose-Einstein condensation in
quantum magnets. \emph{Rev. Mod. Phys.} \textbf{2014}, \emph{86},
563--614. \url{https://doi.org/10.1103/REVMODPHYS.86.563}.

(3) Matsumoto, M.; Normand, B.; Rice, T. M.; Sigrist, M. Field- and
pressure-induced magnetic quantum phase transitions in
TlCuCl\textsubscript{3}. \emph{Phys. Rev. B} \textbf{2004}, \emph{69},
054423. \url{https://doi.org/10.1103/PhysRevB.69.054423}.

(4) Sebastian, S. E.; Harrison, N.; Batista, C. D.; Balicas, L.; Jaime,
M.; Sharma, P. A.; Kawashima, N.; Fisher, I. R. Dimensional reduction at
a quantum critical point. \emph{Nature} \textbf{2006}, \emph{441},
617--620. \url{https://doi.org/10.1038/nature04732}.

(5) Nikuni, T.; Oshikawa, M.; Oosawa, A.; Tanaka, H. Bose-Einstein
condensation of dilute magnons in TlCuCl\textsubscript{3}. \emph{Phys.
	Rev. Lett.} \textbf{2000}, \emph{84}, 5868.
\url{https://doi.org/10.1103/PhysRevLett.84.5868}.

(6) Kodama, K.; Takigawa, M.; Horvatić, M.; Berthier, C.; Kageyama, H.;
Ueda, Y.; Miyahara, S.; Becca, F.; Mila, F. Magnetic superstructure in
the two-dimensional quantum antiferromagnet
SrCu\textsubscript{2}(BO\textsubscript{3})\textsubscript{2}.
\emph{Science} \textbf{2002}, \emph{298}, 395--399.
\url{https://doi.org/10.1126/SCIENCE.1075045}.

(7) R\"uegg, C.; Cavadini, N.; Furrer, A.; G\"udel, H. U.; Kr\"amer, K.;
Mutka, H.; Wildes, A.; Habicht, K.; Vorderwisch, P. Bose--Einstein
condensation of the triplet states in the magnetic insulator
TlCuCl\textsubscript{3}. \emph{Nature} \textbf{2003}, \emph{423},
62--65. \url{https://doi.org/10.1038/nature01617}.

(8) Jaime, M.; Correa, V. F.; Harrison, N.; Batista, C. D.; Kawashima,
N.; Kazuma, Y.; Jorge, G. A.; Stern, R.; Heinmaa, I.; Zvyagin, S. A.;
Sasago, Y.; Uchinokura, K. Magnetic-field-induced condensation of
triplons in Han Purple pigment
BaCuSi\textsubscript{2}O\textsubscript{6}. \emph{Phys. Rev. Lett.}
\textbf{2004}, \emph{93}, 087203.
\url{https://doi.org/10.1103/PHYSREVLETT.93.087203}.

(9) R\"uegg, C.; McMorrow, D. F.; Normand, B.; R\o{}nnow, H. M.; Sebastian,
S. E.; Fisher, I. R.; Batista, C. D.; Gvasaliya, S. N.; Niedermayer, C.;
Stahn, J. Multiple magnon modes and consequences for the Bose-Einstein
condensed phase in BaCuSi\textsubscript{2}O\textsubscript{6}.
\emph{Phys. Rev. Lett.} \textbf{2007}, \emph{98}, 017202.
\url{https://doi.org/10.1103/PHYSREVLETT.98.017202}.

(10) Sengupta, P.; Batista, C. D. Field-induced supersolid phase in
spin-one Heisenberg models. \emph{Phys. Rev. Lett.} \textbf{2007},
\emph{98}, 227201. \url{https://doi.org/10.1103/PHYSREVLETT.98.227201}.

(11) Stone, M. B.; Lumsden, M. D.; Chang, S.; Samulon, E. C.; Batista,
C. D.; Fisher, I. R. Singlet-triplet dispersion reveals additional
frustration in the triangular-lattice dimer compound
Ba\textsubscript{3}Mn\textsubscript{2}O\textsubscript{8}. \emph{Phys.
	Rev. Lett.} \textbf{2008}, \emph{100}, 237201.
\url{https://doi.org/10.1103/PHYSREVLETT.100.237201}.

(12) Samulon, E. C.; Jo, Y. J.; Sengupta, P.; Batista, C. D.; Jaime, M.;
Balicas, L.; Fisher, I. R. Ordered magnetic phases of the frustrated
spin-dimer compound
Ba\textsubscript{3}Mn\textsubscript{2}O\textsubscript{8}. \emph{Phys.
	Rev. B} \textbf{2008}, \emph{77}, 214441.
\url{https://doi.org/10.1103/PHYSREVB.77.214441}.

(13) Samulon, E. C.; Kohama, Y.; McDonald, R. D.; Shapiro, M. C.;
Al-Hassanieh, K. A.; Batista, C. D.; Jaime, M.; Fisher, I. R. Asymmetric
quintuplet condensation in the frustrated S=1 spin dimer compound
Ba\textsubscript{3}Mn\textsubscript{2}O\textsubscript{8}. \emph{Phys.
	Rev. Lett.} \textbf{2009}, \emph{103}, 047202.
\url{https://doi.org/10.1103/PHYSREVLETT.103.047202}.

(14) Oosawa, A.; Fujisawa, M.; Osakabe, T.; Kakurai, K.; Tanaka, H.
Neutron diffraction study of the pressure-induced magnetic ordering in
the spin gap system TlCuCl\textsubscript{3}. \emph{J. Phys. Soc. Jpn.}
\textbf{2003}, \emph{72}, 1026--1029.
\url{https://doi.org/10.1143/JPSJ.72.1026}.

(15) Tanaka, H.; Goto, K.; Fujisawa, M.; Ono, T.; Uwatoko, Y. Magnetic
ordering under high pressure in the quantum spin system
TlCuCl\textsubscript{3}. \emph{Physica B: Condens. Matter}
\textbf{2003}, \emph{329--333}, 697--698.
\url{https://doi.org/10.1016/S0921-4526(02)02009-4}.

(16) R\"uegg, C.; Furrer, A.; Sheptyakov, D.; Str\"assle, T.; Kr\"amer, K. W.;
G\"udel, H. U.; M\'{e}l\'{e}si, L. Pressure-induced quantum phase transition in
the spin-liquid TlCuCl\textsubscript{3}. \emph{Phys. Rev. Lett.}
\textbf{2004}, \emph{93}, 257201.
\url{https://doi.org/10.1103/PHYSREVLETT.93.257201}.

(17) R\"uegg, C.; Normand, B.; Matsumoto, M.; Furrer, A.; McMorrow, D. F.;
Kr\"amer, K. W.; G\"udel, H. U.; Gvasaliya, S. N.; Mutka, H.; Boehm, M.
Quantum magnets under pressure: controlling elementary excitations in
TlCuCl\textsubscript{3}. \emph{Phys. Rev. Lett.} \textbf{2008},
\emph{100}, 205701.
\url{https://doi.org/10.1103/PHYSREVLETT.100.205701}.

(18) Merchant, P.; Normand, B.; Kr\"amer, K. W.; Boehm, M.; McMorrow, D.
F.; R\"uegg, C. Quantum and classical criticality in a dimerized quantum
antiferromagnet. \emph{Nat. Phys.} \textbf{2014}, \emph{10}, 373--379.
\url{https://doi.org/10.1038/nphys2902}.

(19) Oosawa, A.; Ono, T.; Tanaka, H. Impurity-induced antiferromagnetic
ordering in the spin gap system TlCuCl\textsubscript{3}. \emph{Phys.
	Rev. B} \textbf{2002}, \emph{66}, 020405.
\url{https://doi.org/10.1103/PhysRevB.66.020405}.

(20) Manna, S.; Majumder, S.; De, S. K. Tuning of the spin gap
transition of spin dimer
compound~Ba\textsubscript{3}Mn\textsubscript{2}O\textsubscript{8} by
doping with La and V. \emph{J. Phys.: Condens. Matter} \textbf{2009},
\emph{21}, 236005. \url{https://doi.org/10.1088/0953-8984/21/23/236005}.

(21) Rice, T. M. Quantum mechanics: To condense or not to condense.
\emph{Science} \textbf{2002}, \emph{298}, 760--761.
\url{https://doi.org/10.1126/SCIENCE.1078819}.

(22) Sasago, Y.; Uchinokura, K.; Zheludev, A.; Shirane, G.
Temperature-dependent spin gap and singlet ground state in
BaCuSi\textsubscript{2}O\textsubscript{6}. \emph{Phys. Rev. B}
\textbf{1997}, \emph{55}, 8357.
\url{https://doi.org/10.1103/PhysRevB.55.8357}.

(23) Uchida, M.; Tanaka, H.; Mitamura, H.; Ishikawa, F.; Goto, T.
High-field magnetization process in the
Ba\textsubscript{3}Mn\textsubscript{2}O\textsubscript{8}. \emph{Phys.
	Rev. B} \textbf{2002}, \emph{66}, 054429.
\url{https://doi.org/10.1103/PhysRevB.66.054429}.

(24) Matkovich, V. I., Ed. \emph{Boron and Refractory Borides};
Springer: Berlin, 1977. \url{https://doi.org/10.1007/978-3-642-66620-9}.

(25) Scheifers, J. P.; Zhang, Y.; Fokwa, B. P. T. Boron: Enabling
exciting metal-rich structures and magnetic properties. \emph{Acc. Chem.
	Res.} \textbf{2017}, \emph{50}, 2317--2325.
\url{https://doi.org/10.1021/ACS.ACCOUNTS.7B00268}.

(26) Akopov, G.; Yeung, M. T.; Kaner, R. B. Rediscovering the crystal
chemistry of borides. \emph{Adv. Mater.} \textbf{2017}, \emph{29},
1604506. \url{https://doi.org/10.1002/ADMA.201604506}.

(27) Zhang, S.; Wang, Q.; Kawazoe, Y.; Jena, P. Three-dimensional
metallic boron nitride. \emph{J. Am. Chem. Soc.} \textbf{2013},
\emph{135}, 18216--18221. \url{https://doi.org/10.1021/JA410088Y}.

(28) Mbarki, M.; St. Touzani, R.; Fokwa, B. P. T. Unexpected synergy
between magnetic iron chains and stacked B\textsubscript{6} Rings in
Nb\textsubscript{6}Fe\textsubscript{1-x}Ir\textsubscript{6+x}B\textsubscript{8}.
\emph{Angew. Chem.} \textbf{2014}, \emph{126}, 13390--13393.
\url{https://doi.org/10.1002/ANGE.201406397}.

(29) Shankhari, P.; Bakshi, N. G.; Zhang, Y.; Stekovic, D.; Itkis, M.
E.; Fokwa, B. P. T. A delicate balance between antiferromagnetism and
ferromagnetism: Theoretical and experimental studies of
A\textsubscript{2}MRu\textsubscript{5}B\textsubscript{2} (A=Zr, Hf;
M=Fe, Mn) metal borides. \emph{Chem. - Eur. J.} \textbf{2020},
\emph{26}, 1979--1988. \url{https://doi.org/10.1002/CHEM.201904572}.

(30) Sharma, N.; Zhang, Y.; Mbarki, M.; Scheifers, J. P.; Yubuta, K.;
Kimber, S. A. J.; Fokwa, B. P. T.
Nb\textsubscript{6}Mn\textsubscript{1-x}Ir\textsubscript{6+x}B\textsubscript{8}
(x= 0.25): A ferrimagnetic boride containing planar B\textsubscript{6}
rings interacting with ferromagnetic Mn chains. \emph{J. Phys. Chem. C}
\textbf{2021}, \emph{125}, 13635--13640.
\url{https://doi.org/10.1021/ACS.JPCC.1C02662}.

(31) Scheifers, J. P.; Flores, J. H.; Janka, O.; P\"ottgen, R.; Fokwa, B.
P. T. Triangular arrangement of ferromagnetic iron chains in the
high-T\textsubscript{C} ferromagnet
TiFe\textsubscript{1-x}Os\textsubscript{2+x}B\textsubscript{2}.
\emph{Chem. - Eur. J.} \textbf{2022}, \emph{28}, e202201058.
\url{https://doi.org/10.1002/CHEM.202201058}.

(32) Park, H; Encinas, A.; Scheifers, J. P.; Zhang, Y.; Okwa, B. P. T.
F. Boron-dependency of molybdenum boride electrocatalysts for the
hydrogen evolution reaction. \emph{Angew. Chem. Int. Ed.} \textbf{2017},
\emph{56}, 5575--5578. \url{https://doi.org/10.1002/ANIE.201611756}.

(33) Jothi, P. R.; Zhang, Y.; Yubuta, K.; Culver, D. B.; Conley, M.;
Fokwa, B. P. T. Abundant vanadium diboride with graphene-like boron
layers for hydrogen evolution. \emph{ACS Appl. Energy Mater.}
\textbf{2019}, \emph{2}, 176--181.
\url{https://doi.org/10.1021/ACSAEM.8B01615}.

(34) Woo, K. E.; Kong, S.; Chen, W.; Chang, T. H.; Viswanathan, G.;
Díez, A. M.; Sousa, V.; Kolen'ko, Y. V.; Lebedev, O. I.; Costa
Figueiredo, M.; Kovnir, K. Topotactic BI\textsubscript{3}-assisted
borodization: synthesis and electrocatalysis applications of transition
metal borides. \emph{J. Mater. Chem. A} \textbf{2022}, \emph{10},
21738--21749. \url{https://doi.org/10.1039/D2TA04266E}.

(35) Saglik, K.; Mete, B.; Terzi, I.; Candolfi, C.; Aydemir, U.; Saglik,
K.; Mete, B.; Terzi, I. Thermoelectric borides: Review and future
perspectives. \emph{Adv. Phys. Res.} \textbf{2023}, 2300010.
\url{https://doi.org/10.1002/APXR.202300010}.

(36) Yeung, M. T.; Mohammadi, R.; Kaner, R. B. Ultraincompressible,
superhard materials. \emph{Annu. Rev. Mater. Res.} \textbf{2016},
\emph{46}, 465--485.
\url{https://doi.org/10.1146/ANNUREV-MATSCI-070115-032148}.

(37) Akopov, G.; Roh, I.; Sobell, Z. C.; Yeung, M. T.; Pangilinan, L.;
Turner, C. L.; Kaner, R. B. Effects of variable boron concentration on
the properties of superhard tungsten tetraboride. \emph{J. Am. Chem.
	Soc.} \textbf{2017}, \emph{139}, 17120--17127.
\url{https://doi.org/10.1021/JACS.7B08706}.

(38) Akopov, G.; Yeung, M. T.; Roh, I.; Sobell, Z. C.; Yin, H.; Mak, W.
H.; Khan, S. I.; Kaner, R. B. Effects of dodecaboride-forming metals on
the properties of superhard tungsten tetraboride. \emph{Chem. Mater.}
\textbf{2018}, \emph{30}, 3559--3570.
\url{https://doi.org/10.1021/ACS.CHEMMATER.8B01464}.

(39) Akopov, G.; Yeung, M. T.; Sobell, Z. C.; Turner, C. L.; Lin, C. W.;
Kaner, R. B. Superhard mixed transition metal dodecaborides. \emph{Chem.
	Mater.} \textbf{2016}, \emph{28}, 6605--6612.
\url{https://doi.org/10.1021/ACS.CHEMMATER.6B02632}.

(40) Nagamatsu, J.; Nakagawa, N.; Muranaka, T.; Zenitani, Y.; Akimitsu,
J. Superconductivity at 39 K in magnesium diboride. \emph{Nature}
\textbf{2001}, \emph{410}, 63--64.
\url{https://doi.org/10.1038/35065039}.

(41) Kortus, J.; Mazin, I. I.; Belashchenko, K. D.; Antropov, V. P.;
Boyer, L. L. Superconductivity of metallic boron in
MgB\textsubscript{2}. \emph{Phys. Rev. Lett.} \textbf{2001}, \emph{86},
4656. \url{https://doi.org/10.1103/PhysRevLett.86.4656}.

(42) An, J. M.; Pickett, W. E. Superconductivity of
MgB\textsubscript{2}: Covalent bonds driven metallic. \emph{Phys. Rev.
	Lett.} \textbf{2001}, \emph{86}, 4366.
\url{https://doi.org/10.1103/PhysRevLett.86.4366}.

(43) Liu, A. Y.; Mazin, I. I.; Kortus, J. Beyond Eliashberg
superconductivity in MgB\textsubscript{2}: Anharmonicity, two-phonon
scattering, and multiple gaps. \emph{Phys. Rev. Lett.} \textbf{2001},
\emph{87}, 087005. \url{https://doi.org/10.1103/PhysRevLett.87.087005}.

(44) Gutfleisch, O.; Willard, M. A.; Br\"uck, E.; Chen, C. H.; Sankar, S.
G.; Liu, J. P. Magnetic materials and devices for the 21st century:
Stronger, lighter, and more energy efficient. \emph{Adv. Mater.}
\textbf{2011}, \emph{23}, 821--842.
\url{https://doi.org/10.1002/ADMA.201002180}.

(45) Simonson, J. W.; Poon, S. J. Applying an electron counting rule to
screen prospective thermoelectric alloys: The thermoelectric properties
of YCrB\textsubscript{4} and
Er\textsubscript{3}CrB\textsubscript{7}-type phases. \emph{J. Alloys.
	Compd.} \textbf{2010}, \emph{504}, 265--272.
\url{https://doi.org/10.1016/J.JALLCOM.2010.05.110}.

(46) Flipo, S.; Rosner, H.; Bobnar, M.; Kvashnina, K. O.; Leithe-Jasper,
A.; Gumeniuk, R. Thermoelectricity and electronic properties of
Y\textsubscript{1-x}Ce\textsubscript{x}CrB\textsubscript{4}. \emph{Phys.
	Rev. B} \textbf{2021}, \emph{103}, 195121.
\url{https://doi.org/10.1103/PHYSREVB.103.195121}.

(47) Candan, A.; Surucu, G.; Gencer, A. Electronic, mechanical and
lattice dynamical properties of YXB\textsubscript{4} (X = Cr, Mn, Fe,
and Co) compounds. \emph{Phys. Scr.} \textbf{2019}, \emph{94}, 125710.
\url{https://doi.org/10.1088/1402-4896/AB473E}.

(48) Akopov, G.; Pangilinan, L. E.; Mohammadi, R.; Kaner, R. B.
Perspective: Superhard metal borides: A look forward. \emph{APL Mater.}
\textbf{2018}, \emph{6}, 70901. \url{https://doi.org/10.1063/1.5040763}.

(49) Akopov, G.; Yin, H.; Roh, I.; Pangilinan, L. E.; Kaner, R. B.
Investigation of hardness of ternary borides of the
YCrB\textsubscript{4}, Y\textsubscript{2}ReB\textsubscript{6},
Y\textsubscript{3}ReB\textsubscript{7}, and
YMo\textsubscript{3}B\textsubscript{7} structural types. \emph{Chem.
	Mater.} \textbf{2018}, \emph{30}, 6494--6502.
\url{https://doi.org/10.1021/ACS.CHEMMATER.8B03008}.

(50) Kuzma, Y. B. Crystal structure of the YCrB\textsubscript{4}
compound and its analogs. \emph{Sov. Phys. Crystallogr.} \textbf{1970},
\emph{15}, 372--374.

(51) Kuzma, Y. B. New ternary compounds with the structure of
YCrB\textsubscript{4} type. \emph{Dopov. Akad. Nauk Ukr. RSR Ser. A}
\textbf{1970}, \emph{8}, 756--758.

(52) Rogl, P. Chapter 6.7.2.3 Borides with two-dimensional boron
networks. \emph{Inorganic Reactions and Methods} \textbf{1991},
\emph{13}, 156--167. \url{https://doi.org/10.1002/9780470145289.CH26}.

(53) Rogl, P. Chapter 49 Phase equilibria in ternary and higher order
systems with rare earth elements and boron. \emph{Handbook on the
	Physics and Chemistry of Rare Earths} \textbf{1984}, \emph{6}, 335--523.
\url{https://doi.org/10.1016/S0168-1273(84)06006-2}.

(54) Rogl, P. New ternary borides with YCrB\textsubscript{4}-type
structure. \emph{Mater. Res. Bull.} \textbf{1978}, \emph{13}, 519--523.
\url{https://doi.org/10.1016/0025-5408(78)90160-5}.

(55) Medvedeva, N. I.; Medvedeva, Y. E.; Ivanovskii, A. L. Electronic
structure of ternary boron-containing phases YCrB\textsubscript{4},
Y\textsubscript{2}ReB\textsubscript{6}, and
MgC\textsubscript{2}B\textsubscript{2}. \emph{Dokl. Phys. Chem.}
\textbf{2002}, \emph{383}, 75--77.
\url{https://doi.org/10.1023/A:1014734530591}.

(56) Sun, Y.; Zhang, Z.; Porter, A. P.; Kovnir, K.; Ho, K. M.; Antropov,
V. Prediction of Van Hove singularity systems in ternary borides.
\emph{npj Comput. Mater.} \textbf{2023}, \emph{9}, 1--11.
\url{https://doi.org/10.1038/s41524-023-01156-8}.

(57) Jain, A.; Ong, S. P.; Hautier, G.; Chen, W.; Richards, W. D.;
Dacek, S.; Cholia, S.; Gunter, D.; Skinner, D.; Ceder, G.; Persson, K.
A. Commentary: The materials project: A materials genome approach to
accelerating materials innovation. \emph{APL Mater.} \textbf{2013},
\emph{1}, 011002. \url{https://doi.org/10.1063/1.4812323}.

(58) Kirklin, S.; Saal, J. E.; Meredig, B.; Thompson, A.; Doak, J. W.;
Aykol, M.; R\"uhl, S.; Wolverton, C. The Open Quantum Materials Database
(OQMD): assessing the accuracy of DFT formation energies. \emph{npj
	Comput. Mater.} \textbf{2015}, \emph{1}, 1--15.
\url{https://doi.org/10.1038/npjcompumats.2015.10}.

(59) Sun, W.; Dacek, S. T.; Ong, S. P.; Hautier, G.; Jain, A.; Richards,
W. D.; Gamst, A. C.; Persson, K. A.; Ceder, G. The thermodynamic scale
of inorganic crystalline metastability. \emph{Sci. Adv.} \textbf{2016},
\emph{2}, e1600225. \url{https://doi.org/10.1126/SCIADV.1600225}.

(60) Liechtenstein, A. I.; Katsnelson, M. I.; Antropov, V. P.; Gubanov,
V. A. Local spin density functional approach to the theory of exchange
interactions in ferromagnetic metals and alloys. \emph{J. Magn. Magn.
	Mater.} \textbf{1987}, \emph{67}, 65--74.
\url{https://doi.org/10.1016/0304-8853(87)90721-9}.

(61) Pashov, D.; Acharya, S.; Lambrecht, W. R. L.; Jackson, J.;
Belashchenko, K. D.; Chantis, A.; Jamet, F.; van Schilfgaarde, M.
Questaal: A package of electronic structure methods based on the linear
muffin-tin orbital technique. \emph{Comput. Phys. Commun.}
\textbf{2020}, \emph{249}, 107065.
\url{https://doi.org/10.1016/J.CPC.2019.107065}.

(62) Kato, T.; Oosawa, A.; Takatsu, K.; Tanaka, H.; Shiramura, W.;
Nakajima, K.; Kakurai, K. Magnetic excitations in the spin gap system
KCuCl\textsubscript{3} and TlCuCl\textsubscript{3}. \emph{J. Phys. Chem.
	Solids} \textbf{1999}, \emph{60}, 1125--1128.
\url{https://doi.org/10.1016/S0022-3697(99)00072-4}.

(63) Zhu, M.; Matsumoto, M.; Stone, M. B.; Dun, Z. L.; Zhou, H. D.;
Hong, T.; Zou, T.; Mahanti, S. D.; Ke, X. Amplitude modes in
three-dimensional spin dimers away from quantum critical point.
\emph{Phys. Rev. Res.} \textbf{2019}, \emph{1}, 033111.
\url{https://doi.org/10.1103/PHYSREVRESEARCH.1.033111}.

(64) Tanaka, H.; Takatsu, K. I.; Shiramura, W.; Ono, T. Singlet ground
state and excitation gap in the double spin chain system
KCuCl\textsubscript{3}. \emph{J. Phys. Soc. Jpn.} \textbf{1996},
\emph{65}, 1945--1948. \url{https://doi.org/10.1143/JPSJ.65.1945}.

(65) Vinnik, O.; Tarasenko, R.; Orend\'{a}\v{c}, M.; Orend\'{a}\v{c}ov\'{a}, A. Dimerized
nature of magnetic interactions in the S = 1/2 quantum antiferromagnet
Cu(en)\textsubscript{2}SO\textsubscript{4}. \emph{J. Magn. Magn. Mater.}
\textbf{2022}, \emph{547}, 168789.
\url{https://doi.org/10.1016/J.JMMM.2021.168789}.

(66) Carlin, R. L.; van Duyneveldt, A. J. Dimers and clusters. In
\emph{Magnetic Properties of Transition Metal Compounds}; Springer,
Berlin, Heidelberg, 1977; pp 77--108.
\url{https://doi.org/10.1007/978-3-642-87392-8_4}.

(67) Tokuda, M.; Yubuta, K.; Shishido, T.; Sugiyama, K. Redetermination
of the crystal structure of yttrium chromium tetraboride,
YCrB\textsubscript{4}, from single-crystal X-ray diffraction data.
\emph{Acta Crystallogr. E: Crystallogr. Commun.} \textbf{2023},
\emph{79}, 1072--1075. \url{https://doi.org/10.1107/S2056989023008952}.

(68) Bl\"ochl, P. E. Projector augmented-wave method. \emph{Phys. Rev. B}
\textbf{1994}, \emph{50}, 17953.
\url{https://doi.org/10.1103/PhysRevB.50.17953}.

(69) Kresse, G.; Furthm\"uller, J. Efficient iterative schemes for
\emph{ab initio} total-energy calculations using a plane-wave basis set.
\emph{Phys. Rev. B} \textbf{1996}, \emph{54}, 11169.
\url{https://doi.org/10.1103/PhysRevB.54.11169}.

(70) Perdew, J. P.; Burke, K.; Ernzerhof, M. Generalized gradient
approximation made simple. \emph{Phys. Rev. Lett.} \textbf{1996},
\emph{77}, 3865. \url{https://doi.org/10.1103/PhysRevLett.77.3865}.

(71) Samolyuk, G. D.; Antropov, V. P. Character of magnetic
instabilities in CaFe\textsubscript{2}As\textsubscript{2}. \emph{Phys.
	Rev. B} \textbf{2009}, \emph{79}, 052505.
\url{https://doi.org/10.1103/PHYSREVB.79.052505}.

\end{document}